%% file: paper.tex
\documentclass[manuscript,11pt,screen]{acmart}

\usepackage[english]{babel}
\usepackage{subfigure,hyphenat,enumitem}
\usepackage{algorithm}
\usepackage[noEnd]{algpseudocodex}
\usepackage{amsmath,hyperref}
\usepackage{tabularx,tablefootnote}
\usepackage[labelformat=simple,skip=0pt]{subcaption}

\usepackage{tabularx,wrapfig}
\usepackage[capitalize]{cleveref}

\newcommand{\parabreak}{\vspace*{1.00ex minus 0.25ex}\noindent}

\renewcommand{\algorithmicrequire}{\textbf{Input:}}
\renewcommand{\algorithmicensure}{\textbf{Output:}}

\def \bralloc {\mathit{b}_{\mathit{alloc}}}
\def \brspare {\mathit{b}_{\mathit{spare}}}
\def \brvideo {\mathit{b}_{\mathit{video}}}
\def \fbw {\mathit{f}_{\mathit{bw}}}
\def \ffps {\mathit{f}_{\mathit{fps}}}
\def \fres {\mathit{f}_{\mathit{res}}}
\def \rstep {\mathit{r}_{\mathit{step}}}

\renewcommand\footnotetextcopyrightpermission[1]{} % removes footnote with conference info
\setcopyright{none}
\newcommand{\toolname}{NG-Scope5G}
\newcommand{\toolnames}{NG-Scope5G's}
\newcommand{\sysname}{\textsf{TGaming}}
\newcommand{\sysnames}{\textsf{TGaming's}}

\newcommand{\srsNet}{{\small\textsf{[srsRAN/Open5GS]}}}
\newcommand{\aethNet}{{\small\textsf{[Mosolabs/Aether]}}}

\acmDOI{}

\acmISBN{}

\acmConference[]{Paper \#32}
\acmYear{12 pages, plus references}
\copyrightyear{}

\acmPrice{}

\crefdefaultlabelformat{\bfseries#2#1#3}

\DeclareCaptionLabelSeparator{emdash}{--- }

\setitemize{itemsep=2pt,topsep=3pt,parsep=2pt,partopsep=0pt,leftmargin=1.5em}
\setenumerate{itemsep=2pt,topsep=3pt,parsep=2pt,partopsep=0pt,leftmargin=1.5em}
\setlist{itemsep=2pt,parsep=2pt}

\title{Evolving Mobile Cloud Gaming with 5G~Standalone Network Telemetry}

\author{Haoran Wan}
\orcid{0000-0001-9726-195X}
\affiliation{%
  \institution{Department of Computer Science, Princeton University}
  \streetaddress{35 Olden Street}
  \city{Princeton} 
  \state{New Jersey} 
  \postcode{08544}
}
\email{hw8161@princeton.edu}

\author{Kyle Jamieson}
\orcid{0000-0002-7940-2867}
\affiliation{%
  \institution{Department of Computer Science, Princeton University}
  \streetaddress{35 Olden Street}
  \city{Princeton} 
  \state{New Jersey} 
  \postcode{08544}
}
\email{kylej@princeton.edu}

\renewcommand{\shortauthors}{}

\begin{document}
\begin{abstract}
\input{abstract}
\end{abstract}

\maketitle

\thispagestyle{empty}

\input{sections/intro}
\input{sections/related}
\input{sections/design}
\input{sections/impl}

\input{sections/eval}

\input{sections/concl}

\section*{Acknowledgements}

This material is based upon work supported by the National
Science Foundation under Grant Nos. AST-2232457, CNS-2223556 and CNS-2027647.
We gratefully acknowledge technical support from the Aether Onramp team
to deploy our 5G Standalone cell.

\bibliographystyle{concise2}
\begin{raggedright}
\bibliography{paws-zotero}
\end{raggedright}
\appendix
\input{sections/appendix}
\end{document}

%% file: abstract.tex
Mobile cloud gaming places the simultaneous demands of high capacity and low latency on the wireless network, demands that Private and Metropolitan-Area Standalone 5G networks are poised to meet.  However, lacking introspection into the 5G Radio Access Network (RAN), cloud gaming servers are ill-poised to cope with the vagaries of the wireless last hop to a mobile client, while 5G network operators run mostly closed networks, limiting their potential for co-design with the wider internet and user applications.  This paper presents \emph{\toolname{}}, a passive, incrementally-deployable, and independently-deployable Standalone 5G network telemetry system that streams fine-grained RAN capacity, latency, and retransmission information to application servers to enable better millisecond scale, application-level decisions on offered load and bit rate adaptation than end-to-end latency measurements or end-to-end packet losses currently permit.  We design, implement, and evaluate \emph{\sysname{}}, a telemetry-enhanced game streaming platform, demonstrating exact congestion-control that can better adapt game video bitrate while simultaneously controlling end-to-end latency, thus maximizing game quality of experience.  Our experimental evaluation on a production 5G Standalone network demonstrates a 178--249\% Quality of Experience improvement versus two state-of-the-art cloud gaming applications.
%yielding a quality of experience improvement up to  249.2\%. 
% $5\times$ throughput gain, an up to $2\times$ latency reduction and

%% file: sections/intro.tex
\section{Introduction}
\label{s:intro}

\begin{figure*}[tb]
  \centering
  \includegraphics[width=\textwidth]{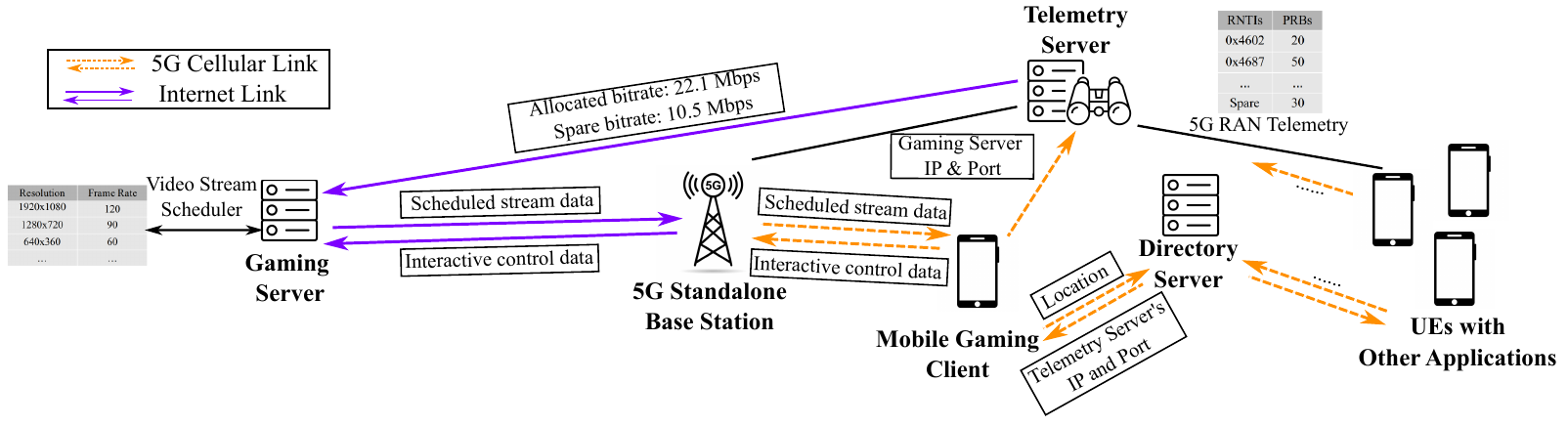}
  \caption{\emph{\sysname{}: System Overview.} \sysnames{} \textbf{Cloud Gaming Server} 
  receives exact capacity estimates of the 5G Standalone Radio Access Network from 
  \sysnames{} \textbf{5G Standalone Telemetry Server}, thus adjusting its video 
  resolution and frame rate 
  to accommodate the RAN's capacity. While the scope of this paper is mobile cloud gaming, 
  our telemetry tool is general, estimating spare capacity for any end\hyp{}to\hyp{}end
  transport protocol or application.}
  \label{f:system_overview}
\end{figure*}

Mobile cloud gaming is a growing business: Amazon (Lu\-na), 
Microsoft (Xbox Game Pass), and Nvidia (GeForce Now) 
all now offer services that stream games 
from the cloud to any desktop, laptop, tablet, mobile phone, or TV device, and 
Apple opening up the App Store to 
game streaming services \cite{webster_apple_2024}.  The promise of a 
high\hyp{}quality gaming experience without the need for expensive, custom hardware 
reflects the current trajectory of industry investment and consumer interest, 
with a majority thinking game streaming services will eventually replace
expensive home consoles \cite{rinderud_cloud_2024}.

The caveat to this pull from the market is the challenge of ensuring 
a reliable and high\hyp{}throughput 
last wireless hop to the mobile gaming device.  With the smartphone 
being the preferred game streaming device,
and the cellular network (Mobile 4G/5G, or 5G \emph{Fixed Wireless Access} 
\cite{nokia_fixed_2023})
the last hop of about half of all game\hyp{}streaming sessions \cite{rinderud_cloud_2024},
it is no surprise that players who stop cloud gaming 
do so primarily because they
perceive poor network quality \cite{rinderud_cloud_2024}.  
High capacity, low latency, 
and high reliability of the last hop are all critical for a seamless user 
experience \cite{rinderud_cloud_2024, hamadanian_ekho_2023}.

Given the above status quo, the evolution of a reliable, low\hyp{}latency,
high\hyp{}capacity 5G cellular 
network is key to ensuring success in the game streaming vertical, and many
others \cite{peterson_5g_2020}.
Game streaming, and other new interactive applications, such as 
augmented reality and remote 
surgery, are intrinsically different from conventional 
video streaming, \emph{e.g.}, because content server\hyp{}to\hyp{}eyeball
latency must be low \cite{cuervo_beyond_2017, krishnan_video_2013},
prohibiting buffering.
5G networks began operating in a \emph{Non-Standalone} 
(NSA) mode with the 2018 3GPP Release~15 \cite{3gpp_release_2018}, which
uses the 4G Evolved Packet Core, which limits performance.  
\emph{5G Standalone} (SA) was introduced in the 2020 Release~16
\cite{3gpp_release_2020}, present in 2022 Release~17 \cite{3gpp_release_2022},
and is central to the ongoing phase of 5G deployment, 
\emph{5G Advanced} (Release~18).  It targets precisely the 
characteristics needed for game streaming: high throughput and low latency
\cite{qualcomm_its_2023}.
In parallel, many organizations are deploying 
\emph{Private 5G} networks 
\cite{redhat_what_2023, peterson_private_2022}---\emph{e.g.}, 
Linux Foundation Aether (formerly ONF Aether)
\cite{onf_onf_2023}---to realize the performance benefits of Standalone
mode at organizational scales. 5G Standalone is thus poised to become 
one of the main bearers for the last hop of
mobile cloud gaming \cite{rinderud_cloud_2024}. 

Yet even in this vibrant commercial landscape, the fundamental challenge 
that remains for cloud game platform designers is to cope with vagaries 
of the wireless Radio Access Network (RAN), 
where capacity changes drastically at millisecond\hyp{}level scales due to
fast fading associated with user mobility and cross\hyp{}traffic from other users
\cite{xie_pbe-cc_2020}.
The game streaming server generates and sends video frames in real\hyp{}time,
in response to game play, which means that game play will stall 
if RAN capacity suddenly falls---the server must reduce its sending 
rate aggressively and instantly.
Transport layer congestion control algorithms such as BBR \cite{cardwell_bbr_2016},
Copa \cite{arun_copa_2018},
and CUBIC \cite{ha_cubic_2008}, video\hyp{}on\hyp{}demand applications 
\cite{fouladi_salsify_2018, mao_neural_2017, akhtar_oboe_2018},
and live and interactive video streaming applications
\cite{kim_neural-enhanced_2020, wang_salientvr_2022}
generally rely on packet loss rates and one\hyp{}way
or round\hyp{}trip latency measurements---a coarse\hyp{}grained stream
of information---to make their bit rate 
adjustment decisions.  This can aggravate queuing in the RAN,
which is also often the end\hyp{}to\hyp{}end bottleneck 
\cite{lee_perceive_2020, xie_accelerating_2017}.

Network telemetry of the RAN can address this challenge,
passively listening to and parsing the operations of the 
RAN to provide granular physical\hyp{}layer capacity 
measurements every millisecond, for every RAN user, without causing 
any interference to the network.  Applications and transport layer protcols
may then adjust the bit rate of their offered load and schedule their
end\hyp{}to\hyp{}end flows more exactly, based on exact estimates of
spare capacity, akin to XCP \cite{katabi_congestion_2002}.

\parabreak{}In this paper, we design and implement \textbf{\sysname{}}, 
a system that enhances mobile cloud gaming performance through 
telemetry for 5G Standalone (SA)
operation. \sysname{} comprises two main components:
\toolname{}, a \emph{5G Standalone Telemetry Server} 
co\hyp{}deployed with the RAN, and video frame scheduler 
and bit rate adaptation logic deployed at the \emph{\sysname{} Gaming Server}.
%The mobile gaming client and gaming server use RTSP 
%\cite{schulzrinne_rfc_1998, schulzrinne_rfc_2016}
%to establish connection and schedule the streams in the session.
%Apart from the normal streams (video, sound for downlink and control for 
%uplink), the gaming server opens another uplink stream for receiving 
%feedback from the telemetry server.
%During the initial RTSP handshake, the client also tells \sysname{} 
%its RNTI and the IP and port of the gaming server so \sysname{} knows
%where to send the feedback.
We design a rendezvous protocol that allows a mobile gaming client
to discover a nearby \sysname{} telemetry server and prompt it to
begin streaming telemetry data over the backhaul network (5G core and wide
area internet), to the gaming server, eschewing the addition of any
telemetry overhead to the wireless medium itself.
Our telemetry server calculates each user's
fair share of any spare capacity, sending this information to the gaming 
server. After receiving \toolname{} telemetry, the \toolname{}\hyp{}enabled
gaming server adjusts its video frame scheduler, and encoding
strategy to match the available downstream capacity.

Importantly, our design is \emph{independently\hyp{}deployable:}
it requires no coordination nor cooperation with
\textbf{1)}~5G mobile network operators, \textbf{2)}~mobile device
manufacturers (both with respect to hardware and software), nor 
\textbf{3)}~5G cellular chipset manufacturers.
Our experimental results show that \sysname{} can 
estimate network flow throughput with a 75th percentile 
error of less than 0.1\% and
detect individual data transmissions with an overall 
miss rate of 0.14\% and 0.22\% for downlink and uplink
respectively.
Our end-to-end evaluation of the \sysname{} cloud gaming platform demonstrates
improved quality of experience (QoE) up to 249\% 
compared with two state-of-the-art game streaming systems.

\parabreak{}Our key contributions are summarized as follows:
\begin{enumerate}
    \item We present the first (to our knowledge) Standalone-5G passive telemetry tool (\emph{\toolname{}}) that can decode some RRC messages and all user control channel information for each user in a 5G SA RAN, at millisecond granularity. 
    
    \item We use \toolnames{} telemetry to design 
    an exact bit rate adaptation algorithm for cloud gaming.
    We design telemetry feedback within our target application's RTSP session, 
    decoupling it from other streams and thus providing a reusable 
    infrastructure service for any future end\hyp{}to\hyp{}end 
    application or transport layer protocol.
    
    \item We evaluate \sysname{} on an open\hyp{}source SA gNB (srsRAN gNB~\cite{system_srsran_2023}\fshyp{}\href{https://open5gs.org}{Open5GS} core) and 
    a commercial 5G SA gNB (Mosolabs \emph{Canopy} small 
    cell \cite{mosolab_mosolab_2023}\fshyp{}Aether 
    Onramp~\cite{onf_onf_2023} core) to demonstrate efficacy and broad
    applicability.
\end{enumerate}

%% file: sections/related.tex
\section{Related Work}
\label{s:related}

\paragraph{Mobile cloud gaming} 
Sunshine \cite{lizardbyte_sunshine_2023} is an open-source C++ cloud 
gaming server. Moonlight \cite{gutman_moonlight_2023} is an open\hyp{}source,
multi\hyp{}platform, NvStreams\hyp{}based \cite{nvidia_nvstreams_2023}
cloud gaming client.  Neither Sunshine nor Moonlight 
contain cellular specific performance optimizations.
Nebula \cite{alhilal_nebula_2022} is an end\hyp{}to\hyp{}end
cloud gaming framework with a similar goal as the present work, which
we compare against head\hyp{}to\hyp{}head in our performance evaluation.
ZGaming \cite{wu_zgaming_2023} uses content\hyp{}specific images to make 
frame predictions and reduce latency,
(an approach orthogonal to \sysname{}),
but needs to train a per\hyp{}game basis.

\paragraph{Interactive video}
Salsify \cite{fouladi_salsify_2018} 
tightly integrates its video codec with a network transport protocol, while
Sprout \cite{winstein_stochastic_2013} uses packet arrival times at the 
receiver to adjust sending rates, but neither explores the use of
RAN telemetry.  Octopus \cite{chen_octopus_2023} 
intentionally overestimates the server's bit\hyp{}rate, then drops 
packets at the 5G RAN, but requires
coordination with the RAN designer and mobile 
operator, unlike \sysname{}.

\paragraph{On-Demand video streaming.} 
MTP \cite{yin_control-theoretic_2015} develops a control theoretic 
adaptive bit rate (ABR) algorithm by solving the quality of 
experience maximization problem. Pensieve \cite{mao_neural_2017} 
generates ABR algorithms using reinforcement learning.
Both rely on previous video chunks' bit rates for their 
model inputs, which is slower and thus poorly\hyp{}matched
to game streaming.
Oboe \cite{akhtar_oboe_2018} tunes ABR parameters based on network conditions, 
while Bola \cite{spiteri_bola_2020} leverages Lyapunov optimization 
based on buffer levels at the client, but neither explores fast-
changing cellular networks.
Swift \cite{dasari_swift_2022} employs layered neural video codecs and ML-based 
encoding to achieve fast reaction time.  
Big data ML network prediction algorithms have shown potential
to improve video on demand 
\cite{yan_learning_2020, sun_cs2p_2016}, but such studies aggregate
broadband and wireless last hops, and so do not necessarily directly point
to improvements for game streaming. 
However, we believe the combination of RAN telemetry data with these approaches
has great potential.
None of the foregoing
supports real time video, making them hard to apply
to the problem at hand.

%Telemetry has been proved to be useful for performance enhancement 
%in many aspects, such as congestion control~\cite{xie_pbe-cc_2020, xie_ng-scope_2022}, 

%video conferencing LTE telemetry \cite{xie_ng-scope_2022} 

\paragraph{Cellular telemetry.}
Most existing telemetry tools are specific to 
4G LTE.  On the handset,
MobileInsight \cite{li_mobileinsight_2016} and QXDM
capture a variety of telemetry from different layers in the L1--L3 cellular 
stack, from a single UE's perspective.
In 4G, control information is in plaintext, so ``sniffer'' tools do not have 
to decode the RACH process (\S\ref{s:design:telemetry:rach}), 
but also cannot verify the correctness of the decoded information
(\cite[FALCON]{falkenberg_falcon_2019},
\cite[LTESniffer]{hoang_ltesniffer_2023}, 
and \cite[NG-Scope]{xie_ng-scope_2022}), 
whereas \sysname{} 
can (\S\ref{s:eval:telemetry}).
LTeye \cite{kumar_lte_2014} does not handle MIMO, 
while OWL \cite{bui_owl_2016}
lacks support for carrier aggregation.
%The downlink control channel in 4G occupies the whole bandwidth of the cell and 
%is not configurable, whereas PDCCH in 5G has configurable bandwidth and
%position, which is more flexible.
In contrast, \sysname{} 
targets 5G Standalone, with its more
configurable control and data channels.

Industrial telemetry tools such as 
KeySight WaveJudge \cite{keysight_keysight_2023} and ThinkRF 
for 5G are closed source, generally
require expensive and specialized hardware, and stop short
of integration with end\hyp{}to\hyp{}end internet protocols and
applications, which is the focus of the present work.
%With no need to guess all these parameters in a brute force way, 
%MSG 4 saves us computational time and further enables us 
%to decode all the DCIs within a TTI duration. TODO: compare NG-Scope.

%\cite{rupprecht_breaking_2019}

%As far as we know, 5G-sniffer \cite{ludant_5g_2023} is the only work that decodes the 5G control channel for user activity classification, however, it requires the channel and user information as input, including user's RNTI, PDCCH's frequency position and duration, DCI format and \textit{etc.}, which are nearly impossible for a normal user to provide, thus making this tool far from practical.

%% file: sections/design.tex
\section{Design}
\label{s:design}

In this section, we elaborate on the system design of the \sysname{}
cloud gaming platform, 
which comprises three main components: \textbf{1)}~the \sysname{} Directory 
Server and \textbf{2)}~the \sysname{} Video Stream Scheduler
(both described in \S\ref{s:design:scheduler}), and
\textbf{3)}~the \toolname{} 5G RAN Telemetry Server (described in \S\ref{s:design:telemetry}).

Cloud gaming differs from conventional video streaming due to its 
highly interactive nature:
the server computes and renders
each users' viewport in real time, encodes the rendered views
into compressed video frames, and sends the video 
frames to the mobile client.
The \sysname{} scheduler (\S\ref{s:design:scheduler}) works along with the video codec, which receives fine-grained data rate estimation from the \toolname{} telemetry server, adapting video resolution and frame rate to accommodate the network bandwidth.
In this way, \sysname{} optimizes latency and frame rate, which are essential for improving QoE. 

\toolname{} telemetry server (\S\ref{s:design:telemetry}) 
is deployed physically near each cloud gaming user. 
 It decodes 5G standalone RAN telemetry, 
the calculates both allocated capacity
and an estimate of fair\hyp{}share spare capacity
for each user in the cell.
The telemetry server then sends these quantities 
back, end\hyp{}to\hyp{}end, to the cloud gaming server to convey these estimates in the coming milliseconds. 
Though there is a one-way delay lag in the reception, we later show that this granular time resolution fits particularly well with the real\hyp{}time 
need of cloud gaming, as it allows \sysname{} to avoid building excess queues in the downlink direction.

\begin{figure}[tbh]
  \centering
  \includegraphics[width=0.7\linewidth]{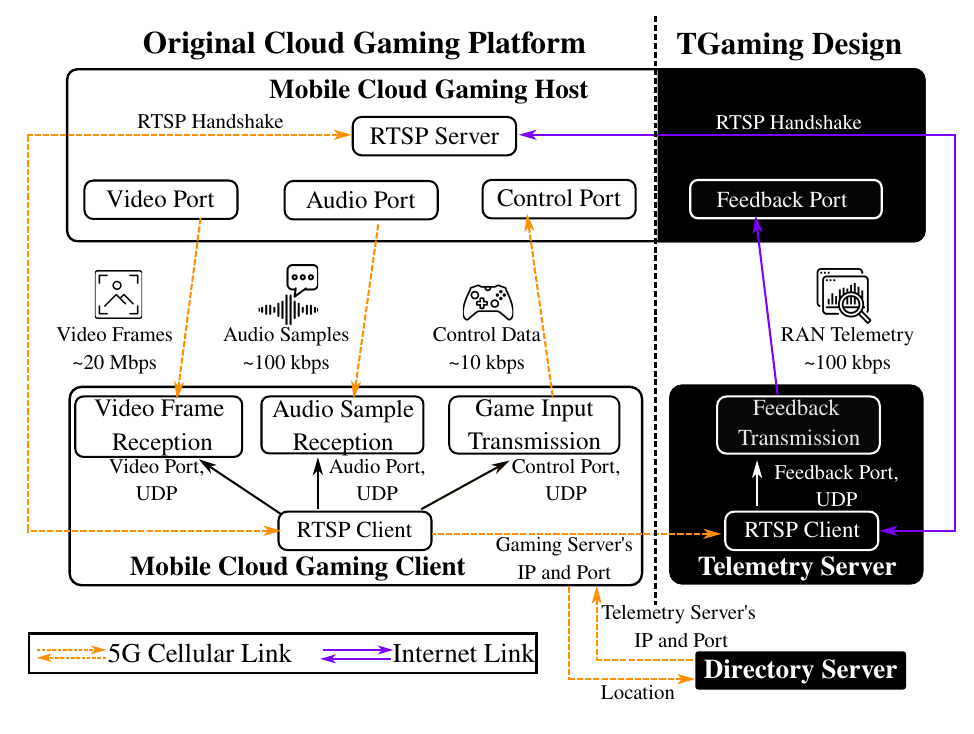}
  \caption{Overview of \sysnames{} overall architecture.\protect\footnotemark}
  \label{fig:gaming-overview}
\end{figure}
\footnotetext{In all figures, we denote \sysnames{} components with
  a black background.}

\input{sections/design_scheduler}
\input{sections/design_telemetry}

%% file: sections/design_scheduler.tex
\subsection{Video Stream Scheduler}
\label{s:design:scheduler}

A mobile cloud gaming server and client set up multiple streams, including video, audio and player control data, using the real\hyp{}time streaming protocol (RTSP) 
\cite{schulzrinne_rfc_1998, schulzrinne_rfc_2016},
an application\hyp{}level protocol for multiplexing and packetizing multimedia transport streams over a suitable transport layer protocol.
Many cloud gaming platforms~\cite{huang_gaminganywhere_2013}, as well as real-time video streaming and teleconferencing platforms use RTSP.
According to our experiments (\S\ref{s:eval}), video data comprises the greatest share of throughput among all cloud gaming streams, thus we focus there on video stream optimization.
\Cref{fig:gaming-overview} shows the overall data flow of \sysnames{} data and control streams, with \sysname{} functionality highlighted.

\subsubsection{Cloud Gaming Session Establishment}
\label{s:design:scheduler:establishment}
We design an RTSP\hyp{}mediated feedback control stream establishment 
protocol for telemetry data that
retains compatiblity with Sunshine, as well as the aforementioned applications. 
In the startup stage, the client makes an RTSP handshake with the mobile cloud gaming server, during which both sides exchange a service description, such as each stream's port and network protocol, agree on parameters, and connect the client to these streams.  
This original handshake and the stream exchange process is represented by the arrow between RTSP server and client in the left\hyp{}hand side of \cref{fig:gaming-overview}.\footnote{Further detail of these RTSP messages exchanged in the handshake is included in Appendix~\ref{appd:rtsp}.}

\paragraph{Rendezvous.} We envision the \toolname{} telemetry server
(at bottom right 
in \cref{fig:gaming-overview}) as
ubquitously and incrementally\hyp{}deployable infrastructure, 
located near clients (and 5G base stations), that measures
the RAN's resource allocation constantly and passively.
Before starting the gaming session, a client queries the 
\sysname{} Directory Server with a rough estimate of its location,
obtained with standard mobile phone OS location services.  The 
Directory Server replies with a list of IP addresses and ports of 
nearby \toolname{} telemetry servers.

\paragraph{Telemetry stream establishment.}
Also during game sesssion establishment, the client software
tells the telemetry server 
the IP address and port of the cloud gaming server, so the 
telemetry server can set up a RAN Telemetry stream (see \cref{fig:gaming-overview})
with the gaming 
server in the same manner the mobile gaming client initiates other 
streams, \textit{e.g.} video, audio, and control.
This design incurs minimal burden on the mobile client and 
5G RAN, because there are no hardware or cellular modem
modifications required for the mobile client.  It also has
the advantage of adding no 
extra data burden to the RAN during the gameplay phase, as the 
RAN telemetry data itself traverses the RAN's (usually wired, 
or distinct and high\hyp{}capacity point to point wireless) backhaul to the 
core network and then over wired links to the mobile cloud gaming server.
With this RTSP initiated feedback design, the telemetry and stream scheduler is 
therefore fully decoupled from the original gaming or streaming operation, 
which provides more flexibility, and the client can decide whether to use it or not. 
During the gameplay phase, the cloud gaming server receives RAN 
telemetry from the telemetry server and uses it to optimize the 
downlink video stream.

\begin{figure}[tbh]
  \centering
  \includegraphics[width=0.6\linewidth]{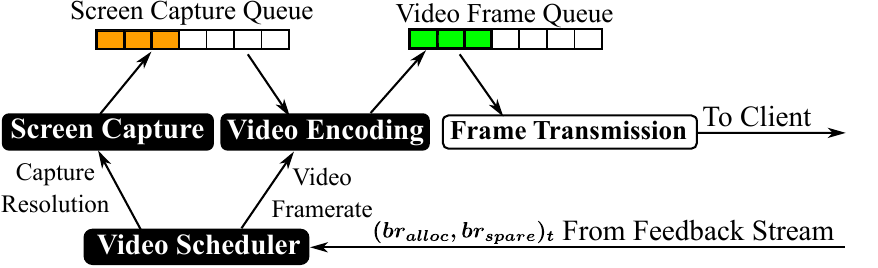}
  \caption{Video capture and scheduling 
  components at the \sysname{} game streaming server.}
  \label{fig:video-scheduler}
\end{figure}

\begin{figure*}[htb]
    \centering   
    \subfigure[When the bottleneck is in the RAN.]{
        \includegraphics[width=0.31\linewidth]{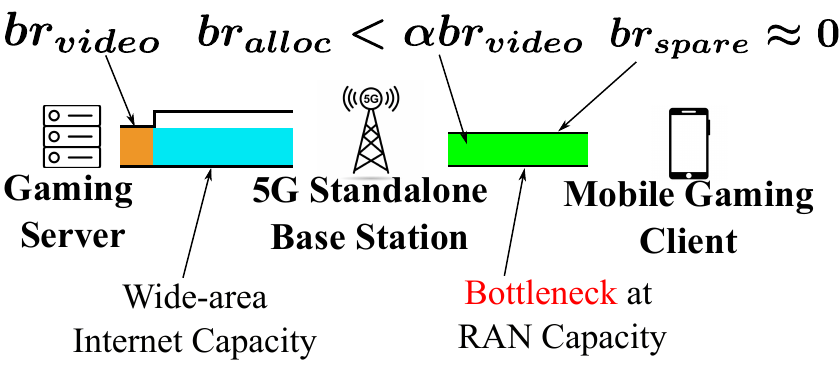}
        \label{fig:bn_ran}}
    \hfill
    \subfigure[When the bottleneck is in the core network.]{
        \includegraphics[width=0.31\linewidth]{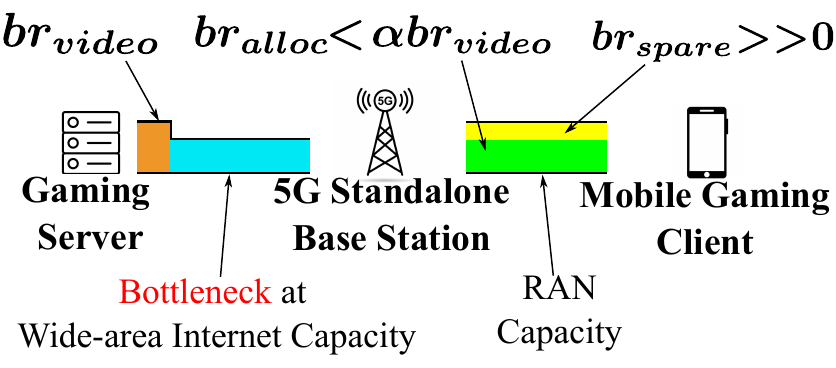}
        \label{fig:bn_core}}
    \hfill  
    \subfigure[When there is not bottleneck.]{
        \includegraphics[width=0.31\linewidth]{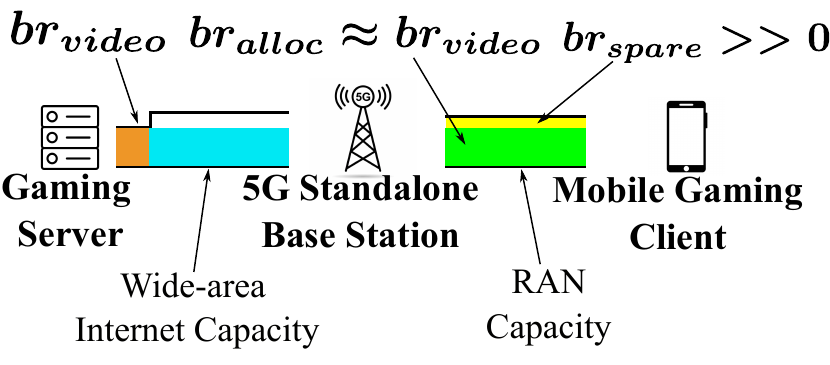}
        \label{fig:bn_none}}
     \caption{The possible situations with bottleneck in different position of the network.} \label{fig:bn_scenarios}
\end{figure*}

\subsubsection{Gameplay}
\label{s:design:stream:gameplay}
During the gameplay phase, the server's video processing consists 
of the three threads of execution shown in \cref{fig:video-scheduler}:
\textsf{screen capture}, 
\textsf{video encoding}, and \textsf{frame transmission}.
Unlike conventional video\hyp{}on\hyp{}demand transmission 
schemes that work on a per\hyp{}\emph{chunk} (a collection of video frames that lasts several seconds) basis 
\cite{yin_control-theoretic_2015}, cloud gaming's video 
capture and transmission work on a per\hyp{}frame basis.
The client receives video frames from the server, decodes them also on a 
per\hyp{}frame basis, and displays the frames on the screen.
As long as the client can handle the video frames in real time, the client's buffer will not grow, and so the frame decoding capability of client sets the cap for the video stream scheduler's rate, which, in our case on a USD~200
phone, is 1080p video at 120 frames\fshyp{}s (fps).

Each element in the telemetry server--gaming server
feedback stream consists 
of a tuple of 32\hyp{}bit 
unsigned integers, indexed by transmission time interval $t$: 
\begin{equation}
\left[\left(\bralloc,\brspare\right)_t\right].
\end{equation}
The elements of this tuple represent the instantaneous (measured at
time $t$) \emph{bit\hyp{}rate allocated to the user} and \emph{spare bit\hyp{}rate
available in the RAN},
in units of bits per second.
%
%
% We will further introduce how to conduct telemetry in RAN and get those values in detail in Section~\ref{sec:telemetry-tool}.
%
When the \textsf{video scheduler} needs to change one of the video stream's parameters
(resolution or frame rate),
it raises a \emph{reinit} event and adjusts the variable(s) corresponding to
the changing parameter(s).
The \textsf{screen capture} and \textsf{video encoding} threads
check the status of this event after processing of each frame, if the \emph{reinit} is raised, the two threads load the new values before they process the next frame.
In this way, the scheduler is able to change the video stream's resolution or frame rate within duration of one frame, \textit{e.g.} $16.67~ms$ for 60 fps video.

During gameplay, the \textsf{video scheduler} actively
compares the allocated and spare bit\hyp{}rates from the feedback stream with the current bit\hyp{}rate of the video stream to make real\hyp{}time, millisecond\hyp{}granularity scheduling decisions.
To calculate the latter, we leverage Moonlight's 
functionality (see \S\ref{s:impl} for further details) 
to map resolution and frame rate into a 
\emph{current video bit rate} $\brvideo$
(we denote this mapping $\fbw$):
\begin{equation}
    \brvideo = \fbw(w, h, r),
\end{equation}
where $w$ and $h$ are current video frames' width and height in number of pixels, and $r$ is current frame rate.
Back\hyp{}computing, given a desired video bit\hyp{}rate $\brvideo$ and a frame resolution $(w,h)$,
we can calculate the frame rate $r$ required to match this bit\hyp{}rate:
we denote this calculation $\ffps$.
Similarly, fixing bit\hyp{}rate and frame rate we calculate a corresponding frame resolution: we denote this calculation
$\fres$.

\paragraph{Bit rate adaptation.}
The \sysname{} \textsf{video scheduler} inputs
allocated and spare RAN bit rates and current video parameters,
computes a target allocated bit rate, and from that, target video parameters
for the next second of time.  

The scheduler first estimates whether
the end\hyp{}to\hyp{}end bottleneck link is located in the RAN, or elsewhere 
in the core or wide\hyp{}area internet.
If $\brvideo$ exceeds $\alpha\cdot\bralloc$ (for a parameter $\alpha>1$), then 
the cloud gaming server is 
sending at a faster rate than the RAN has allocated (as in \cref{fig:bn_ran}).\footnote{We 
choose $\alpha \gets 1.75$ here as a slightly loose bound to desensitize 
\sysname{} to imprecision in Moonlight's bit rate estimation.}
If $\brspare$ 
is also positive, then the end\hyp{}to\hyp{}end 
path's bottleneck link is likely in the core network or wide\hyp{}area internet because
the excess packets are queuing at the bottleneck link and not 
saturating the RAN (as in \cref{fig:bn_core})---in this case, we target an unchanged bit rate.

Otherwise, either the RAN is meeting the gaming server's offered bit rate (as in \cref{fig:bn_none}), 
or it is saturated (possibly by other users as well), or both.  
We can therefore step up the offered bit rate to the extent there is spare
capacity: we add $\bralloc$ and a 75\% of $\brspare$ 
together to arrive at our new target bit rate. 
We can take all the $\brspare$ in the RAN as in previous work \cite{xie_pbe-cc_2020}, but the RAN may have different scheduling decisions so we take a conservative 75\% of $\brspare$.
Summarizing, 
\begin{equation}
\bralloc'=\begin{cases}
    \bralloc & \text{if $\brvideo > \alpha\cdot\bralloc$, $\brspare > 0$}\\
    \bralloc + \frac{3}{4}\brspare & \text{otherwise}.
    \end{cases}
\end{equation}
%Based on $\bralloc'$ and $\brvideo$, the scheduler makes the real-time decisions.

\begin{algorithm}
\caption{\sysnames{} video stream bit rate adaptation.}
\label{alg:video-algo}
\begin{small}
\begin{algorithmic}
\sffamily
    \State \algorithmicrequire\ $br_{alloc}$, $br_{spare}$, $w$, $h$, $r$ 
    \State \algorithmicensure\ $w'$, $h'$, $r'$
    \State $\mathit{fps} \gets [90, 60, 45, 30, 15]$
    \State $\brvideo \gets \fbw (w, h, r)$
    \If{$\brvideo > \alpha \bralloc$}
        \State $\bralloc' \gets \bralloc$
    \Else 
        \State $\bralloc' \gets \bralloc + 0.75\cdot\brspare$
    \EndIf
    \If{$\bralloc' > \brvideo$}
        \If{$w < w_{cap}$ and $h < h_{cap}$}
            \State $w', h' \gets \fres (\bralloc', r)$;
            \State $\left(w',h',r'\right) \gets 
                \left(\max(w', w_{cap}), \max(h', h_{cap}), r\right)$
        \ElsIf{$r < r_{cap}$}
            \State $r' \gets \ffps(\bralloc', w, h)$
            \State $(w',h',r') \gets 
                \left(w, h, \max(r', r_{cap})\right)$
        \EndIf
    \Else
        \State $\rstep \gets r\text{'s position in } \mathit{fps}$;
        \State $w', h' \gets f_{res}(\bralloc', r)$;
        \While{$w' < w_{bad}$ and $h' < h_{bad}$ and $r_{step} < 5$}
            \State $r' \gets \mathit{fps}[r_{step}]$
            \State $w', h' \gets f_{res}(\bralloc', r')$
        \EndWhile
    \EndIf
\end{algorithmic}    
\end{small}
\end{algorithm}

The overall scheduling strategy is shown in 
Algorithm~\ref{alg:video-algo}, where the $w_{cap}$, $h_{cap}$, $r_{cap}$ are 
the upper cap of video's $w$, $h$ and $r$ for the client's decoder.

Given our new target bit rate, we compute new target video parameters.
If $\bralloc'$ is bigger than the current video stream's bandwidth usage, 
$\brvideo$, we first fix the frame rate and increase the video's resolution
until it reaches the resolution cap.
If the video capture's resolution reaches its cap and there is still 
spare bandwidth, we then increase the frame rate for a more responsive experience.
The reasoning behind this is that 
higher video resolution and fixed frame rate provide superior user 
mean opinion scores \cite{jarschel_gaming_2013}.
Conversely, when available bandwidth shrinks, we aggressively decrease 
resolution and frame rate by first reducing the frame rate to
a low preset value, and then gradually decreasing the resolution.

As available channel bandwidth continuously decreases, we first decrease
the frame resolution.
If the resolution drops under a threshold value, 
\textit{e.g.} under $640\times 360$, denoted as $w_{bad}$ and $h_{bad}$, 
we next decrease the frame rate.
The frame rate setting steps are $[90, 60, 45, 30, 15]$~fps.
To achieve high QoE, we avoid changing the resolution at
big steps, \textit{e.g.} from 1080P to 720P.
Instead, we adjust the resolution with smaller steps through $\fres$, such 
as decreasing $10\%$ width and height pixels of 
$1920\times 1080$ video frame into $1728\times 972$ video frame.
The mobile client's video decoder works well with such video resolution 
adjustments, and shrinking the resolution values gradually also helps 
to improve the user's experience because the resolution difference 
is not too obvious.

%% file: sections/design_telemetry.tex
\subsection{A Standalone 5G Telemetry Tool}
\label{s:design:telemetry}

\begin{table*}
    \begin{small}
    \begin{tabularx}{\linewidth}{@{}*2{l}*1{X}@{}}\toprule
    \multicolumn{2}{@{}l}{\textbf{Channel Name}}& \textbf{Description and Relevance}\\\midrule
    PBCH& Physical Broadcast Channel& Signals a cell's existence to UEs and \toolname{}.\\
    PDCCH& Physical Downlink Control Channel& Contains telemetry data locating network traffic and indicating its size.\\
    PDSCH& Physical Downlink Shared Channel& Signals cell parameters to UEs and \toolname{}; 
    delivers user data from the gNB to the UE.\\
    RACH& Random Access Channel& Uplink channel used by the UE to rendezvous with the 5G service.\\
    \bottomrule
    \end{tabularx}
    \end{small}
    \caption{5G control and data channels, in order of their first appearance, 
    and their relevance to \toolname{}.}
    \label{t:5g_channels}
\end{table*}

The high\hyp{}level goal of the \toolname{} telemetry tool is to decode and interpret
\emph{Downlink Control Information} (DCI), fine\hyp{}grained telemetry information 
that the 5G TDD Radio Access Network (RAN)
broadcasts onto the airwaves in the clear and which tells the mobile (referred to
as the UE, for user equipment) where to receive its downlink data from or send its
uplink data to in the physical data channels of the Standalone 5G TDD RAN, and 
thus how much data is actually present on those channels.  
We strive for an open design that operates independently of the 5G network operator and 
5G mobile devices themselves, to decouple our design from any need to coordinate 
with these closed entities.

\paragraph{Preliminaries.}
The RAN divides time and frequency into \emph{Physical Resource Blocks} (PRBs),
equally\hyp{}sized time\hyp{}frequency blocks air that carry information
to or from the base station, also known as a \emph{cell}, or \emph{gNodeB} (gNB).
Time is divided into \emph{slots}, or \emph{Transmission Time Intervals}
(TTIs) which are also the units of downlink\hyp{}uplink transmission switching.
Unlike the 4G LTE RAN, a sub-6 5G RAN can have three different subcarrier spacings 
in the RACH process and later communication, including 15~kHz (same as 4G), 
30, and 60~kHz, which result in TTIs of 1, 0.5,
and 0.25~ms respectively.

Two key channels contain the relevant information: the \emph{Physical Downlink Control 
Channel} (PDCCH), which in general carries control traffic
from the gNB to the network, and the \emph{Physical Downlink Shared Channel} 
(PDSCH), which delivers network traffic from the gNB to the UE.
In general, DCI information in the PDCCH points to PRBs in the PDSCH 
containing both cell configuration information and network traffic.
The \emph{aggregation level} is the number of repetition for one 
DCI's transmission, \textit{e.g.} downlink and uplink DCI uses different
aggregation level in Figure~\ref{fig:dci_overview}, the 
gNB may use a higher aggregation level when channel conditions are poor,
to ensure correct reception of DCI. 

The \emph{Cell Radio Network Temporary Identifier} (C-RNTI) is a 
unique identifier used in
the RAN to identify a specific mobile device or UE.\footnote{The 
\emph{System Information 
Radio Network Temporary Identifier} (SI-RNTI) is the RNTI broadcast address,
fixed as 0xFFFF by the 3GPP standard \cite{3gpp_release_2022}.}
The cell assigns the UE a C-RNTI in an exchange in the 
\emph{Random Access Channel}
(RACH, see \cref{t:5g_channels}) between UE and gNB called the 
\emph{RACH process}, which establishes a Radio Resource Control (RRC) 
session between UE and gNB \cite{takeda_understanding_2020}.
RRC is a Layer-3  protocol whose major functions include connection 
establishment and release functions, and broadcast of system information.

\begin{figure}[tb]
  \centering
  \includegraphics[width=0.6\linewidth]{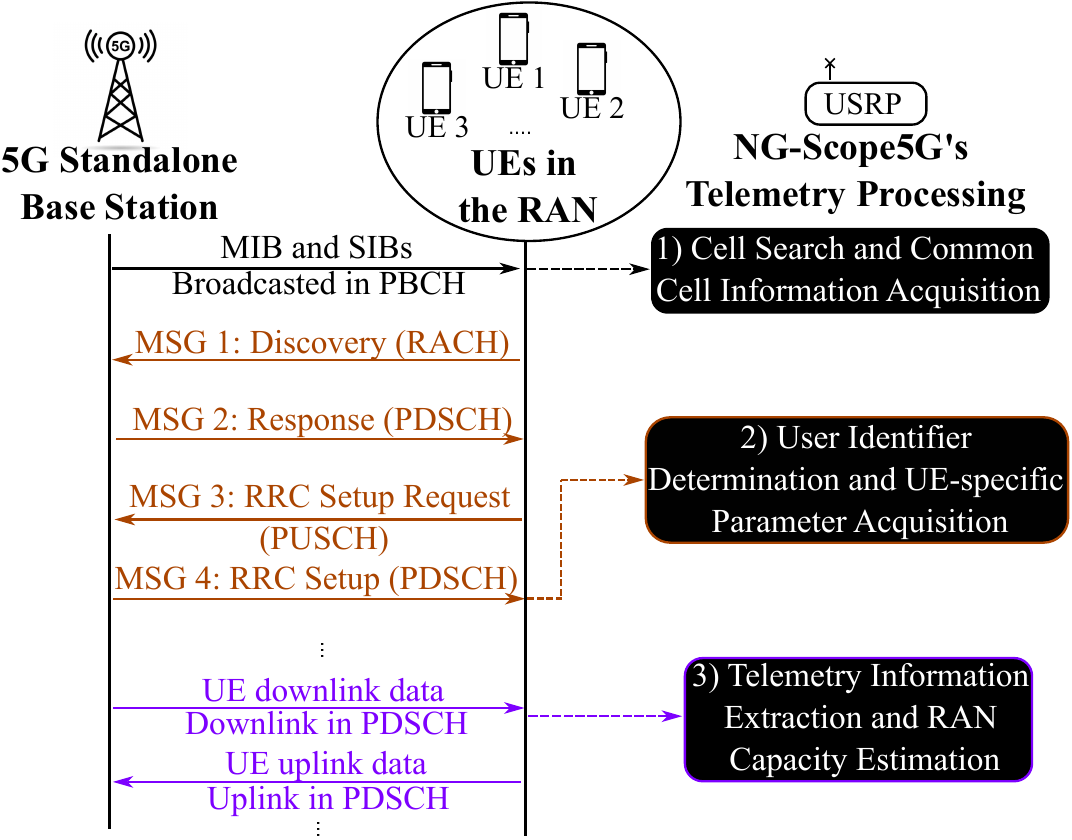}
  \caption{\emph{\toolname{} telemetry design overview:} \toolname{} 
  passively listens to raw radio data and decodes
  radio resource control messages and DCIs from it. \toolnames{} processing 
  components are denoted by a black background.}
  \label{fig:telemetry-overview}
\end{figure}

\parabreak{}An overview of \toolnames{} process is shown 
in Figure~\ref{fig:telemetry-overview}: 
the left side shows standard 5G RAN operation and the physical channels used for 
each message, while the right side shows our processing of telemetry, comprising 
the following high\hyp{}level steps: \textbf{1)}~cell search and common 
(\emph{i.e.}, shared among all UEs) cell parameter acquisition
(\S\ref{s:design:telemetry:cell_search}),
\textbf{2)}~user identifier determination and UE-specific parameter acquisition
(\S\ref{s:design:telemetry:rach}),
\textbf{3)}~telemetry information extraction (\S\ref{s:design:telemetry:dci}),
and \textbf{4)}~RAN capacity estimation (\S\ref{s:design:telemetry:capacity}).

\subsubsection{Cell Search and Cell Parameter Acquisition}
\label{s:design:telemetry:cell_search}

\begin{figure}[tb]
  \centering
  \includegraphics[width=0.7\linewidth]{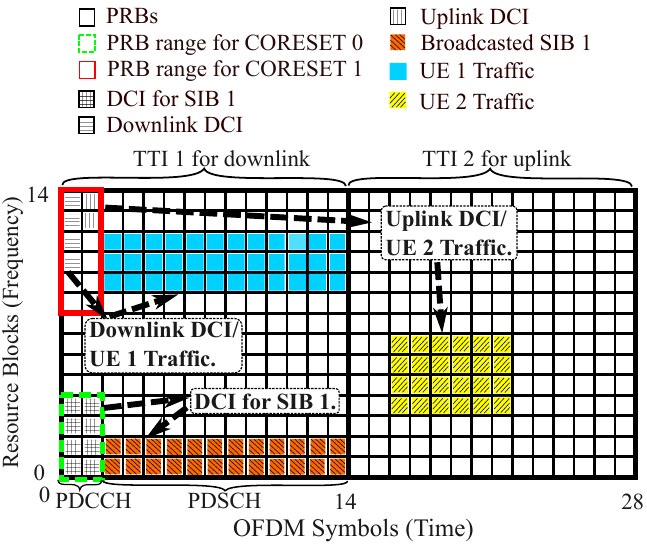}
  \caption{An overview of DCI and data transmission in the 5G physical layer. 
  There are multiple CORESETs, each of which occupies part of the bandwidth, and the 
  DCIs point to the resource blocks that UE should receive from or send data to.}
  \label{fig:dci_overview}
\end{figure}

The goal of this step is to discover 
the cell, and extract its
channel configuration parameters, which describe the structure of 
the channel that the UE later uses to associate with the cell
(\S\ref{s:design:telemetry:rach}).
To make UEs aware of its existence, the base station  
periodically broadcasts basic information about itself in a
\emph{Master Information Block} (MIB), which includes a time index
(the \emph{system frame number}\footnote{One system frame is 10~ms, 
and the system frame number 
(0--1024), which the UE will also need to synchronize, indexes each system frame.}) 
and configuration parameters of initial association.
The MIB is the first message that the UE receives from a base station,
and it is located in the \emph{Physical Broadcast Channel} 
(PBCH, see \cref{t:5g_channels}).
After decoding the MIB, the UE uses the information therein to locate
PRBs that contain the \emph{Control Resource Set} (CORESET)~0,
a part of the larger PDCCH.
CORESET~0 in particular carries a DCI that points to PRBs 
containing the next message we need to decode, the
\emph{System Information Block} (SIB)~1 in
the PDSCH---this is illustrated in the lower\hyp{}left
corner of \cref{fig:dci_overview}.

The SIB~1 (also broadcasted periodically in the
PDSCH channel),
carries common information about the cell, including 
physical channel configuration and
all the information a UE may need for the RACH processing described next
in \S\ref{s:design:telemetry:rach}, 
such as the subcarrier spacing for the cell in RACH process, 
the parameter and time-frequency position for MSG 1 in RACH, 
and aggregation level of the DCIs.
We decode the DCI aggregation level from the SIB~1, obviating the
need to blindly search levels as prior tools do
\cite{xie_ng-scope_2022, falkenberg_falcon_2019}.

%The DCI and SIB 1 reception procedure is similar to the process shown in 
%\cref{fig:dci_overview}, except that the gNB uses the SI-RNTI 
%for broadcasting instead of a UE-specific C-RNTI, and a sample of key
%components in SIB 1 and some extra details can be found in the 
%Appendix~\ref{appd:sib1} ({\color{red} more introduction needed}).

\subsubsection{User Identifier and Parameter Determination}
\label{s:design:telemetry:rach}

The goal of this step is to acquire the user's C-RNTI and the
cell's parameters for the channel  
containing the final telemetry information we seek
(the PDCCH, see \cref{t:5g_channels}).
%The RACH is the first channel used by the UE to access the 5G service in
%the uplink direction
Since DCIs after the RACH process are all scrambled by a sequence 
derived from the C-RNTI, we decode
the parts of the process required to extract the C-RNTI. 

%Furthermore, the DCI format and PDCCH position configuration space in 5G is much larger than that in 4G and only with this information can we make the decoded DCIs into meaningful grants.

There are four messages exchanged between gNB and UE: MSG~1
(Preamble Transmission), MSG~2 (Random Access Response), 
MSG~3 (RRC Setup Request), and MSG~4 (RRC Setup), 
shown on the left side of 
Figure~\ref{fig:telemetry-overview} SIB~1 
tells the UE where and when to transmit MSG~1 to the cell,
to begin the process.\footnote{MSG~1
contains a synchronization sequence that makes 
the cell aware of the UE.
%Both the UE and the gNB use 
%these parameters from MSG~1 to calculate the random access RNTI (RA-RNTI).
%Then the gNB uses the RA-RNTI to send the DCI, which points to
%the resource blocks that contain MSG 2.
MSG 2 contains timing adjustment, a temporary cell RNTI 
(TC-RNTI), 
%the RAPID (Random Access Preamble ID) matching the preamble sent by the UE, 
and an initial uplink grant for the UE to transmit MSG 3.
MSG 3 contains an RRC Setup request to the gNB.
Then the UE and gNB will use the TC-RNTI for MSG 4, after which the UE is RRC connected to the gNB.  
The gNB promotes the TC-RNTI to a C-RNTI, and uses
it for data communication later.}
MSG 4 contains most of the UE-specific information required for mobile communication and for telemetry, such as position and search space of initial CORESET, and the aggregation level for DCI transmission.
\toolname{} only decodes MSG~4, as 
the DCI for MSG~2 and 4 are not scrambled with the RNTI as 
is the case for other DCIs, but instead are in plain text, 
with a TC-RNTI\hyp{}scrambled CRC of the DCI appended, the same as 
in 4G~\cite{3gpp_requirements_2008}.
Knowing this, we can calculate the TC-RNTI by an exclusive\hyp{}or of 
a CRC of the received DCI plain text computed by \toolname{},
and the appended TC-RNTI\hyp{}scrambled CRC of the DCI, separately
received by \toolname{}, as previous 4G sniffers
\cite{xie_ng-scope_2022,falkenberg_falcon_2019} do for
each and every received DCI.
When the TC-RNTI is promoted to the C-RNTI after MSG~4, we thus
obtain the C-RNTI of the UE without decoding the 
preliminary messages and we can verify the correctness of decoded 
MSG 4 through CRC check. 
Then from MSG~4, we get the position of the CORESET, aggregation level of DCIs, 
correct format of DCI that the gNB and UE will use later data communication.
A sample of MSG 4 can be found in Appendix~\ref{appd:msg4}.

\subsubsection{Telemetry Information Extraction}
\label{s:design:telemetry:dci}

The goal of this step is to estimate the amount of 
data traffic in the data channel (PDSCH, see \cref{t:5g_channels}) 
for each UE in the RAN, and its associated physical layer parameters 
(bit rate and channel quality).
%The SIB decoding and RACH process are conducted in CORESET~0 
%and the initial \emph{bandwidth part} (BWP), which is a fraction of the 
%whole cell bandwidth.  After the RACH process, the UE may move to a 
%different CORESET and BWP for communication in the next TTI, 
%informed by MSG~4.
Furthermore, the UE also gets the DCI format type and aggregation 
level from MSG 4 \cite{3gpp_release_2022}\footnote{For example, format 
1-1 or 1-0 for downlink DCI---the sample configuration can be found at \texttt{searchSpacesToAddModList} in Appendix~\ref{appd:msg4}.}
Since we have decoded the required information, 
\toolname{} moves to the same \emph{bandwidth part} (a fraction
of the whole cell channel) as the UE 
for DCI reception at this stage.
With all required information known (C-RNTI, aggregation level, 
DCI format), \toolnames{} DCI reception process is the same as 
the standard 3GPP DCI decoding for each UE, where the detailed
information can be found in~\cite{sharetechnote_pdcch_2023,
takeda_understanding_2020}, which yields the 30--80~bits of DCI 
data.\footnote{This includes deinterleaving and the 
modulation and coding scheme indicator, which are required for transport
block size (TBS) calculation and our later capacity 
estimation. A sample of DCI with format 1-1 and its translated 
downlink grant for PDSCH can be found in Appendix~\ref{appd:dci-and-grant}.}

\subsubsection{RAN Capacity Estimation}
\label{s:design:telemetry:capacity}

The goal of this step is to calculate the capacity (bit rate)
allocated to each UE and the spare RAN capacity, 
through telemetry information.

With the DCI and all the RRC messages we decode from the previous stages, we calculate 
the \emph{Transport Block Size} (TBS) for each DCI, whose value indicates 
how many bits are transmitted in this TTI, for the specific UE through this DCI. 
We calculate the TBS according to the 3GPP standard---it 
is a function of the error control coding rate, the modulation, the number and spacing
of OFDM subcarriers, and framing overheads, whose details we leave 
to Appendix~\ref{appd:tbs}.
We record the TBS for every UE in every TTI, maintain a sliding window to calculate 
the bit rate for each UE, and we update the data rate in every TTI.
Furthermore, in each TTI, we split unused PRBs evenly across UEs and
recalculate these PRBs to yield a fair\hyp{}share
spare capacity attributable to each UE.

%Furthermore, UEs in the RAN may move to locations where the signal strength is weak, such as position 2 and 3 for under blockage status in our experiment (\cref{fig:eval_locations}).
%Base station acquires the channel quality index (CQI) feedback from UE periodically and chooses MCS accordingly \cite{3gpp_release_2022}.
%
%If the channel condition is bad, the gNB decreases the MCS index, which results in a lower bit rate for each PRB but better robustness.
%
%To maintain the same downlink bit rate, the gNB allocates more PRBs to the UE, however, when other UEs kick in, the resource blocks allocated for this UE decreases, which greatly affects the throughput and latency. 
%
Due to wireless fading, the UE may fail to decode the downlink 
traffic, triggering re\hyp{}transmission, in which case
transport blocks contain old data that should not be considered in 
\toolnames{} bit rate calculation.
\toolname{} captures the re\hyp{}transmission information in 
the physical and MAC layers through tracking Hybrid Automatic Repeat Request 
(HARQ) information in DCIs.
The gNB allocates up to 16 HARQ processes for each UE, informing the UE 
of each process through \texttt{harq\_id} in the DCI.
If the UE correctly decodes the data in one TTI, it sends an ACK to the gNB, which
then toggles the \texttt{new\_data\_indicator} (\texttt{ndi}) of the 
DCI with the same \texttt{harq\_id} to indicate 
new data in the next TTI of the process.
If the UE NACKs, the gNB uses the same \texttt{ndi} as the 
previous DCI with the same \texttt{harq\_id} for the re\hyp{}transmission.
\toolname{} maintains a 16\hyp{}element array for each UE to record
the \texttt{ndi} from previous DCIs for each \texttt{harq\_id} to detect
re\hyp{}transmissions, and sets the TBS for this TTI to
zero if it is a re\hyp{}transmission.

% In this circumstance, decreasing the bit rate can help maintain the continuity of the experience and avoid packet loss, so we decrease $(\bralloc, \brspare)_t$ when the telemetry server observes the number of low MCS indexes (smaller than 20) in the past 200 TTIs (100 ms with 30 kHz SCS) is bigger than 50.
% %
% If we select a dynamic value (such as 1/3 of the current bit rate), it may break the video stream transmission because the server will drop the bit rate according to the feedback, which further causes the feedback bit rate to be lower.
% %
% Instead, we select a fixed value -- the bit rate that one PRB can achieve in a low MCS, which can be roughly calculated as the maximum bit rate achieved by all the PRBs divided by the number of PRBs then divided by two (low MCS), it's around 1 Mbits/s in our gNB. 
%
%Through the accurate capacity estimation for each UE, we can adjust the transmission strategy timely in the video scheduler(\S\ref{s:design:scheduler}).

%% file: sections/impl.tex
\section{Implementation}
\label{s:impl}

\begin{figure}[tb]
  \centering
  \includegraphics[width=\linewidth]{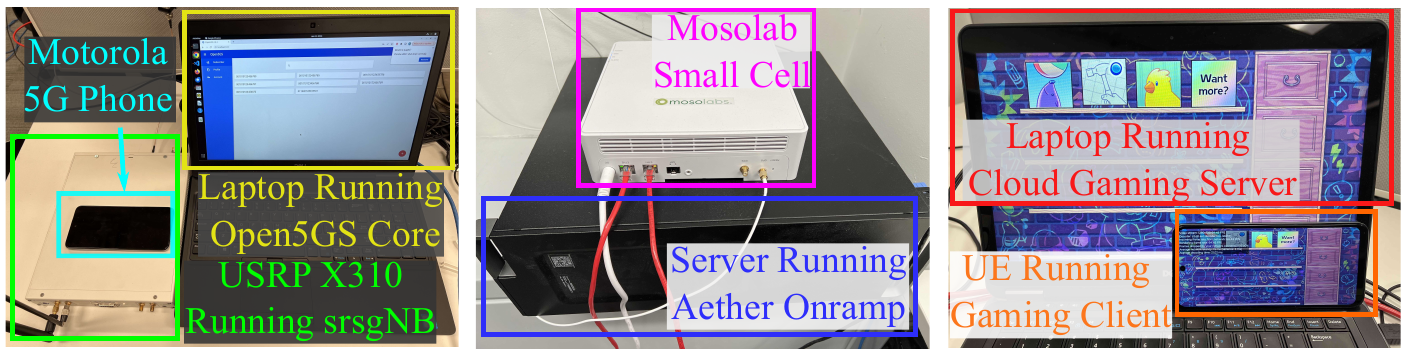}
  \caption{Hardware used in the \sysname{} implementation.}
  \label{fig:hardware}
\end{figure}

We implement \sysnames{} video stream scheduler on top of Sunshine and 
Moonlight with 621 lines of C++ code in the Sunshine server and 224 lines of
Java code in the Moonlight \cite{gutman_moonlight_2023}
android client (excluding reused code).
%Moonlight's core functions are written in C and it has multiple wrappers to support different platforms, such as Android, IOS, Window, Linux and \textit{etc.} 
We extend the event-driven RTSP server to set up the 
feedback stream from \toolnames{} telemetry server. 
The cloud gaming server receives two 32 bits unsigned integer feedback $(br_{alloc}, br_{spare})_t$ from telemetry server in a interval of 0.5 millisecond.
It uses these values to make video scheduling decisions, changing resolution and 
frame rate in real time (\S\ref{s:design:scheduler}).
The client determines the availability of the telemetry server 
and sends the cloud gaming server's address to the telemetry server.

We implement \toolname{}'s telemetry functionality in 3,839 lines of C++ code  
(excluding reused code) processing radio 
signals received by a USRP.
We reuse the physical layer signal processing modules from an open-source 5G library srsRAN 
\cite{system_srsran_2023}, \emph{i.e.}, a wireless channel estimator,
a demodulator, and a frame synchronizer.
srsRAN only supports processing for a 10 MHz FDD base station with 
15 kHz subcarrier spacing, thus we modify their low level processing 
code extensively, altering radio sample reception, Fourier transform 
scheduling and phase compensation, to support different subcarrier 
spacing, higher bandwidth and TDD base station processing, such as 
30 kHz subcarrier spacing, 20 MHz bandwidth in the 5G TDD band. 
Then we implement the cell search, SIB 1 decoding, RACH decoding, DCI 
extraction and RAN throughput estimation on top of that. 
The telemetry server decodes the DCIs within every TTI and calculate the 
allocated bit rate and spare bit rate for each UE in the RAN.
%Furthermore, \sysname{} estimates the physical channel quality through the MCS value in the DCI and adjusts the feedback values $(br_{alloc}, br_{spare})_t$(\S\ref{s:design:telemetry}).

%% file: sections/eval.tex
\begin{figure}
\centering
\includegraphics[width=.7\linewidth]{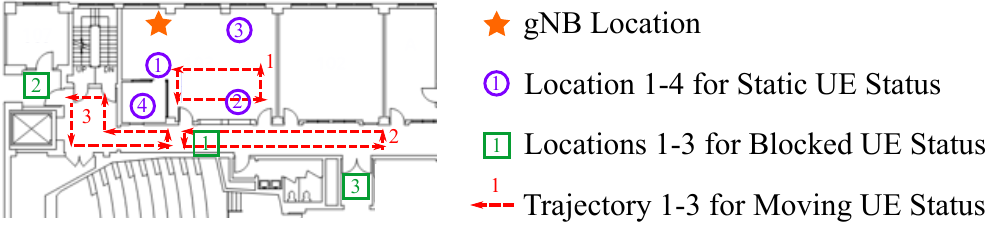}
\caption{Map of performance evaluation.}
\label{fig:eval_locations}
\end{figure}

\section{Evaluation}
\label{s:eval}

We proceed top\hyp{}down, first introducing our overall methodology 
(\S\ref{s:eval:methodology}), then
evaluating end\hyp{}to\hyp{}end cloud gaming performance in terms 
of network utilization and QoE metrics (\S\ref{s:eval:cloudgaming}).
Finally, we
evaluate \toolnames{} telemetry
in terms of DCI decoding, PRB detection and 
flow throughput estimation accuracy (\S\ref{s:eval:telemetry}).

\subsection{Methodology}
\label{s:eval:methodology}

We evaluate \sysname{} in the following two real 5G Standalone networks,
as shown in \cref{fig:hardware}.  In both networks, 
Motorola \emph{Moto G} 5G phones serve as the clients.

\parabreak{}\noindent{}\textbf{\srsNet{}:} 
In this open\hyp{}source 
5G Standalone network, we run \href{https://open5gs.org}{Open5GS} 
on a four\hyp{}core CPU laptop with 8~GB of memory, and an srsRAN gNB 
\cite{system_srsran_2023} with a X310 USRP.
The gNB runs on 5G New Radio band n41 in TDD mode, with downlink center 
frequency 2524.95~MHz, subcarrier spacing 30~kHz, and channel bandwidth 20~MHz.

\parabreak{}\noindent{}\textbf{\aethNet:} In this 
Private 5G Standalone
network, we configure an Aether Onramp 5G 
software defined core network
\cite{onf_onf_2023} on a 10\hyp{}core CPU machine with 16~GB of memory,
and a Sercomm Mosolabs small cell \cite{mosolab_mosolab_2023}.
The gNB runs in the CBRS band (n48) in TDD mode, 
controlled by a 
spectrum access server (SAS) \cite{fcc_35_2023}, on center frequency 
3561.6~MHz, subcarrier spacing 30~kHz, and channel bandwidth 20~MHz.

%We run \sysname{}'s telemetry server on a 16-core machine with 32 GB of memory.

\begin{figure*}[tb]
\centering
    \subfigure[When 5G RAN is less crowded.]{
        \includegraphics[width=0.48\linewidth]{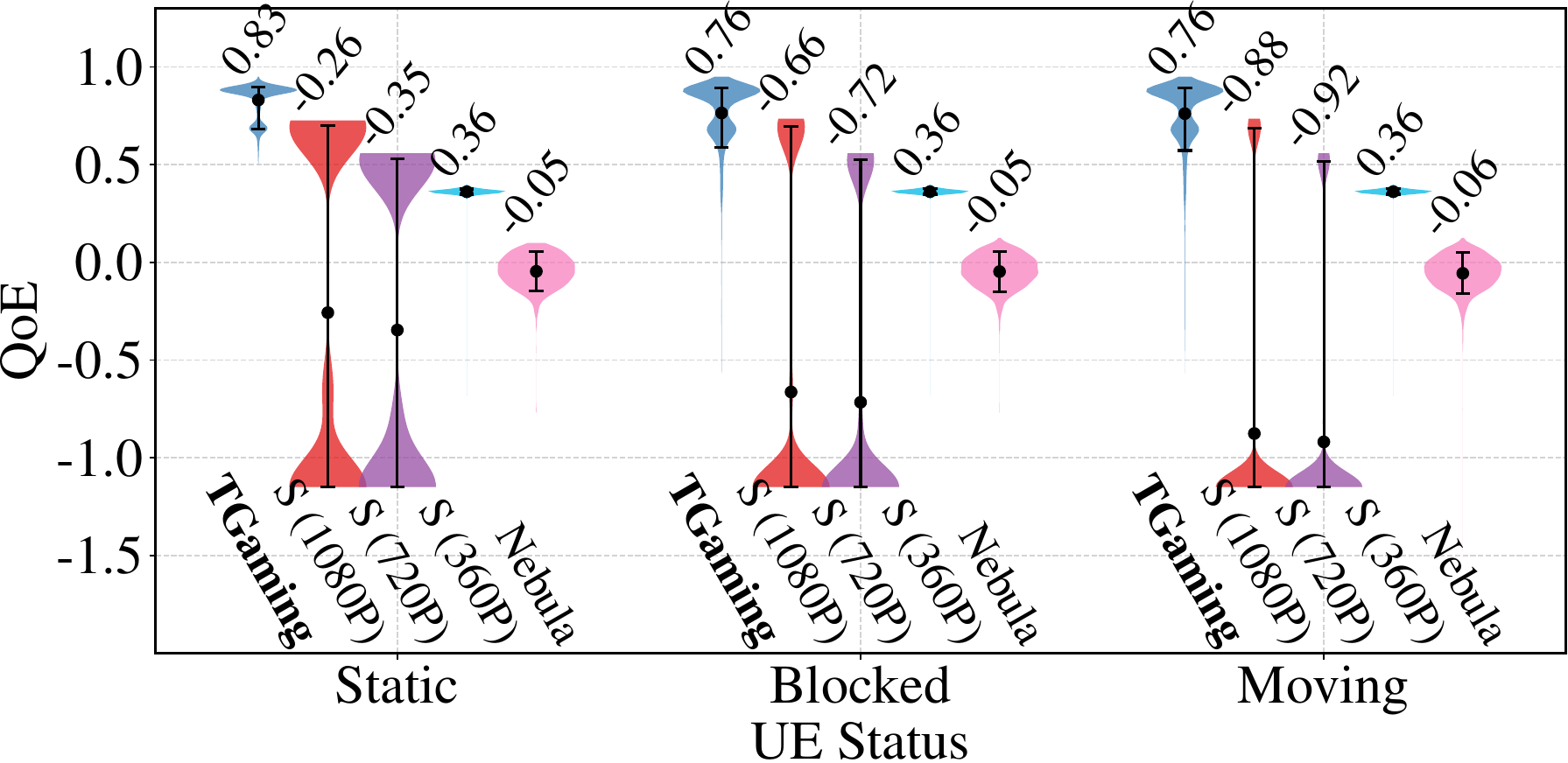}
        \label{fig:qoe_idle}}
        \hfill
    \subfigure[When 5G RAN is crowded.]{
        \includegraphics[width=0.48\linewidth]{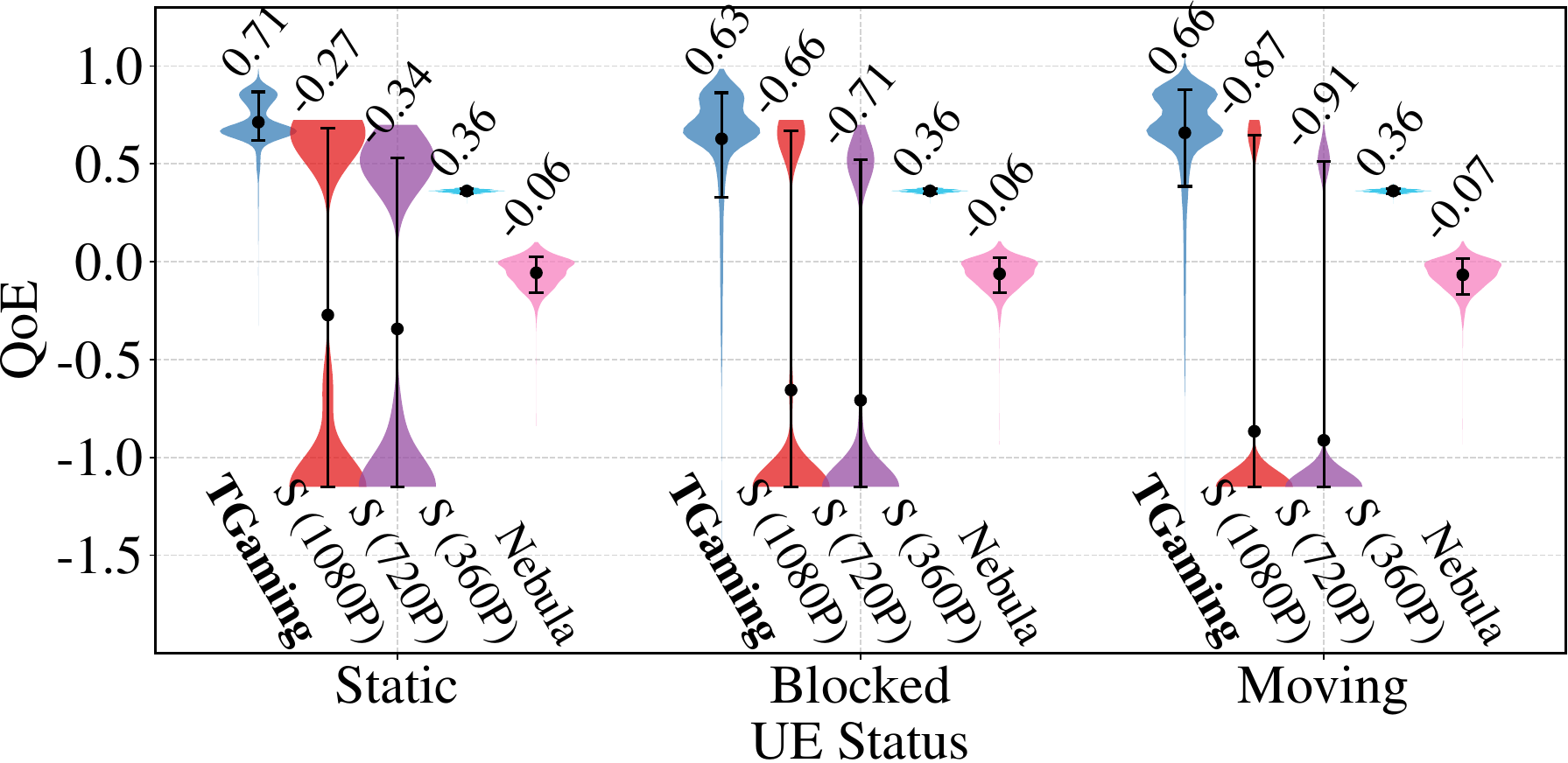}
        \label{fig:qoe_packed}}
     \caption{\emph{QoE comparison:} numbers above data points indicate mean QoE; whiskers denote
     10th\fshyp{}90th percentiles (S: Sunshine).} \label{fig:qoe}
\end{figure*}

\input{sections/eval_gaming}

\subsection{RAN Telemetry}
\label{s:eval:telemetry}

We evaluate \toolname{} tele\-met\-ry time series data, DCI miss rates, 
PRB decoding accuracy, RAN allocated throughput estimation accuracy, and 
wall clock processing time.

\begin{figure*}
\begin{minipage}[b]{1.0\textwidth}
    \centering
    \subfigure[Static, less crowded RAN.]{
        \includegraphics[width=0.313\textwidth]{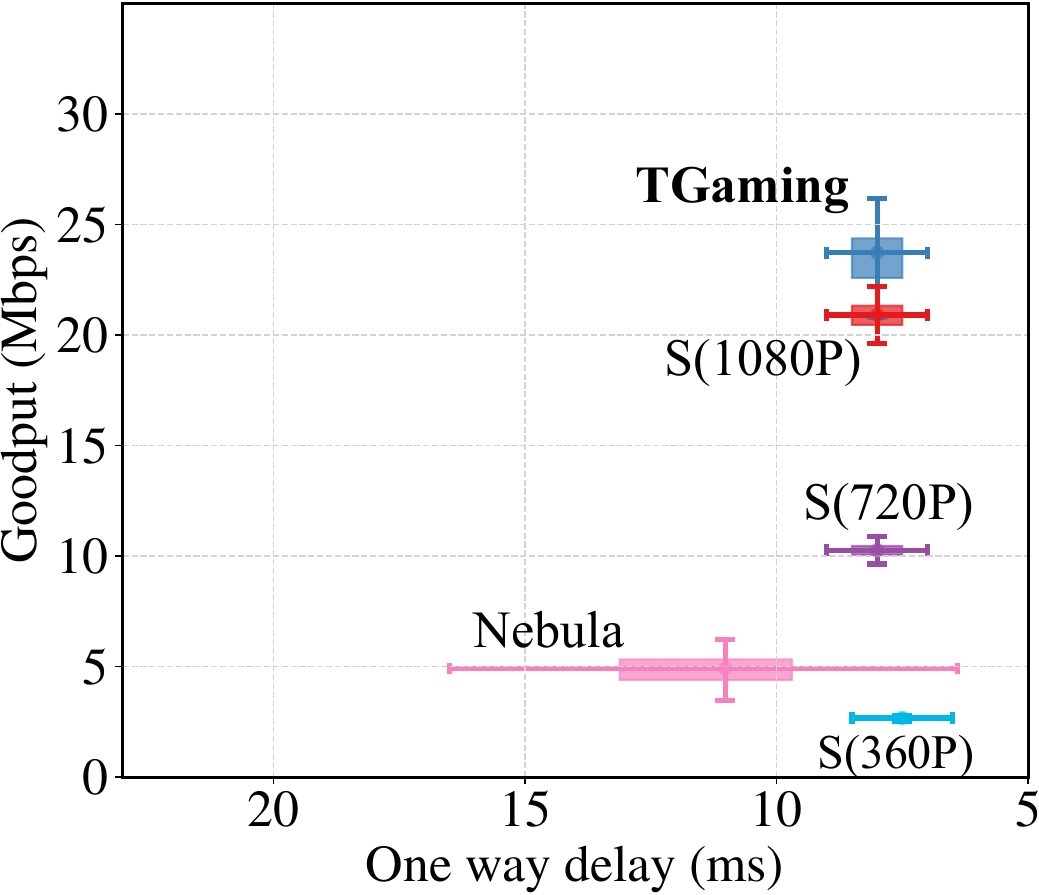}
        \label{fig:bw_idle_static}}
        \hfill
    \subfigure[Under blockage, less crowded RAN.]{
        \includegraphics[width=0.313\textwidth]{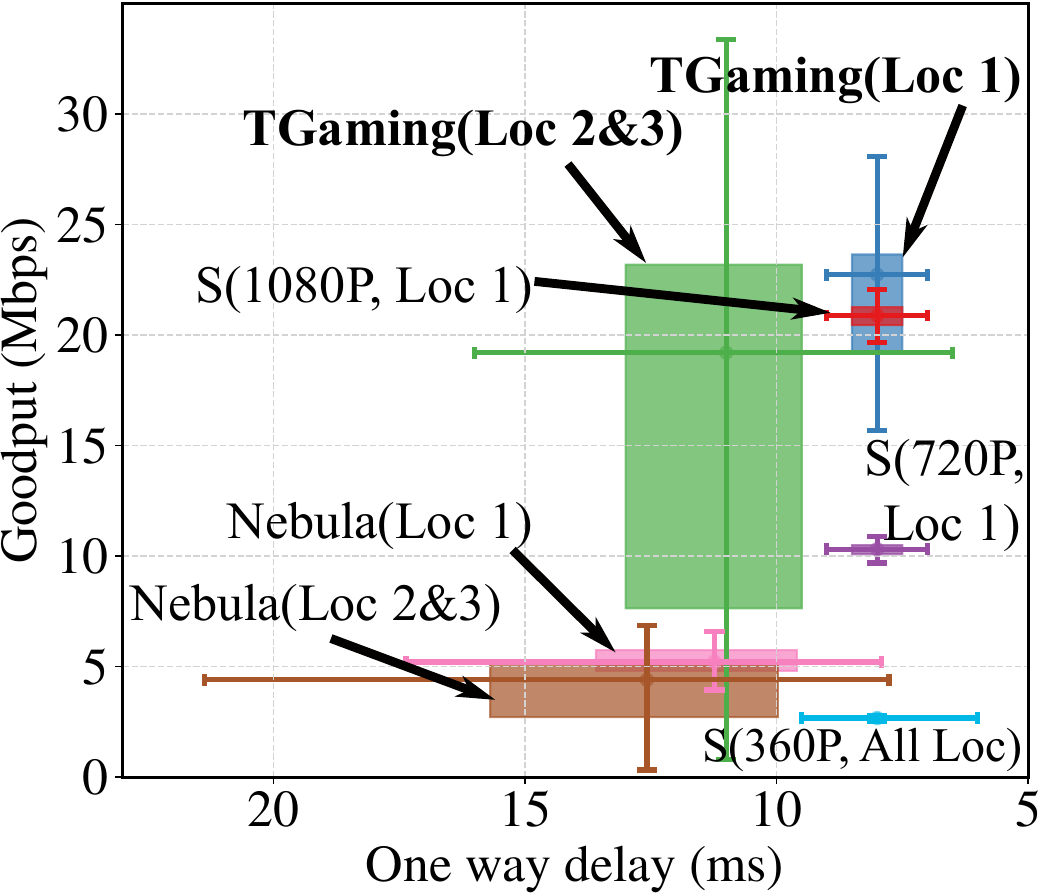}
        \label{fig:bw_idle_blockage}}
    \hfill  
    \subfigure[Moving, less crowded RAN.]{
        \includegraphics[width=0.313\textwidth]{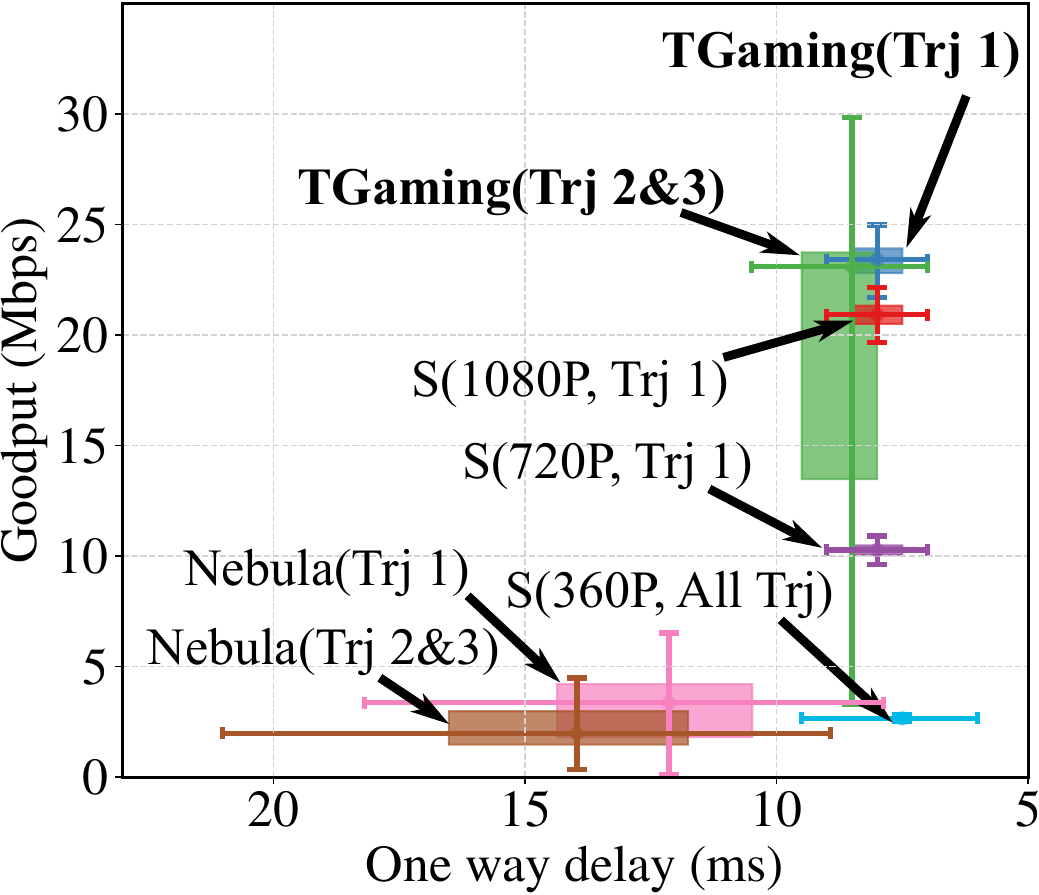}
        \label{fig:bw_idle_moving}}
    \hfill 
    \subfigure[Static, crowded RAN.]{
        \includegraphics[width=0.313\textwidth]{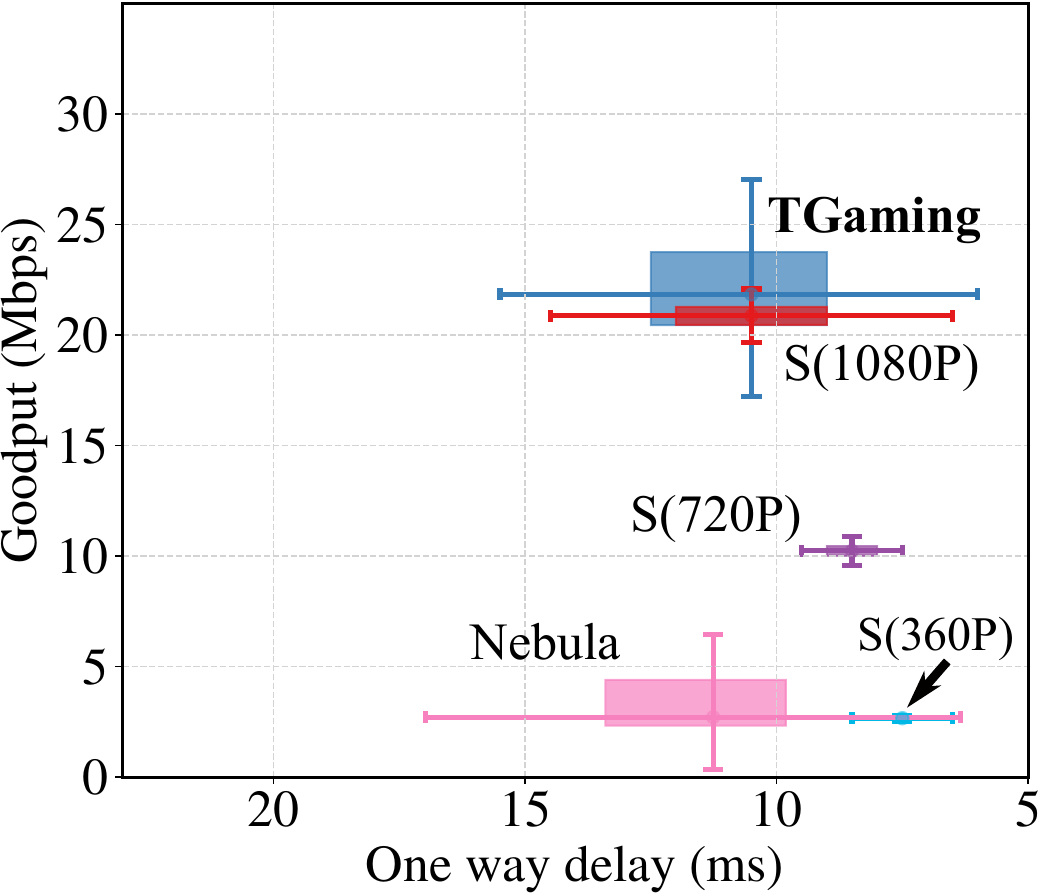}
        \label{fig:bw_packed_static}}
    \hfill
    \subfigure[Under blockage, crowded RAN.]{
        \includegraphics[width=0.313\textwidth]{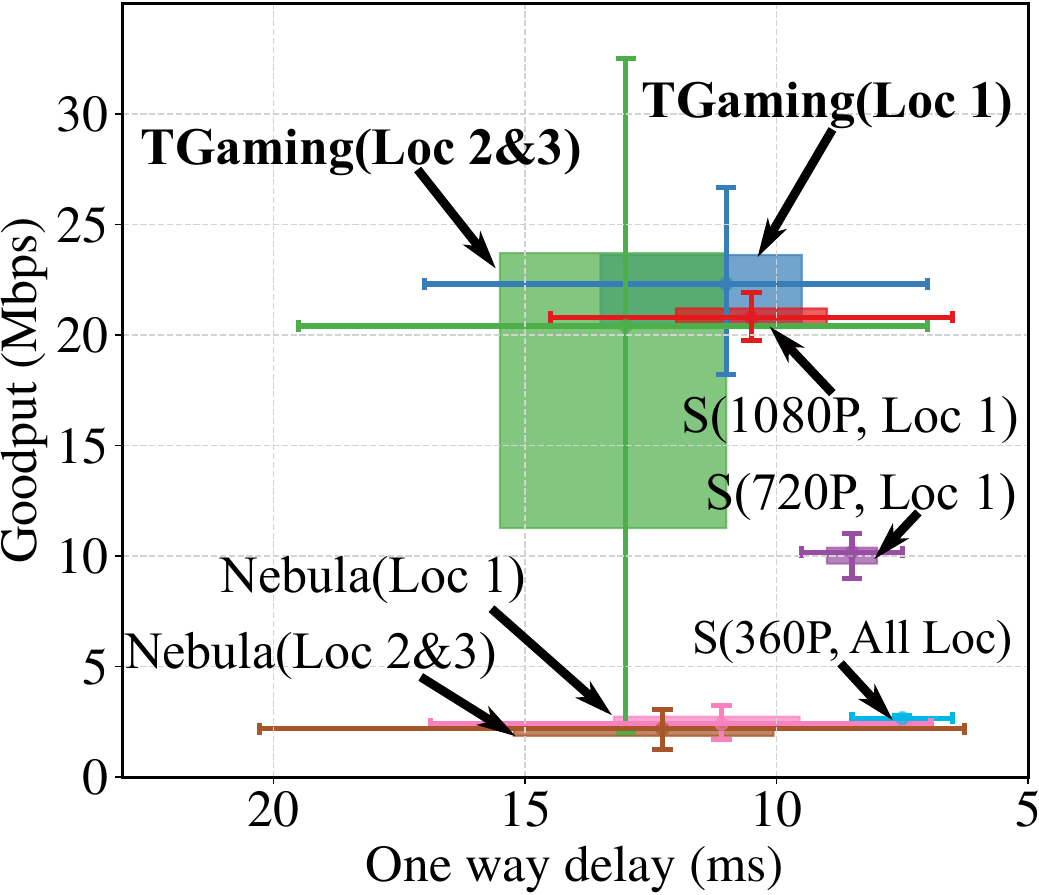}
        \label{fig:bw_packed_blockage}}
    \hfill  
    \subfigure[Moving, crowded RAN.]{
        \includegraphics[width=0.313\textwidth]{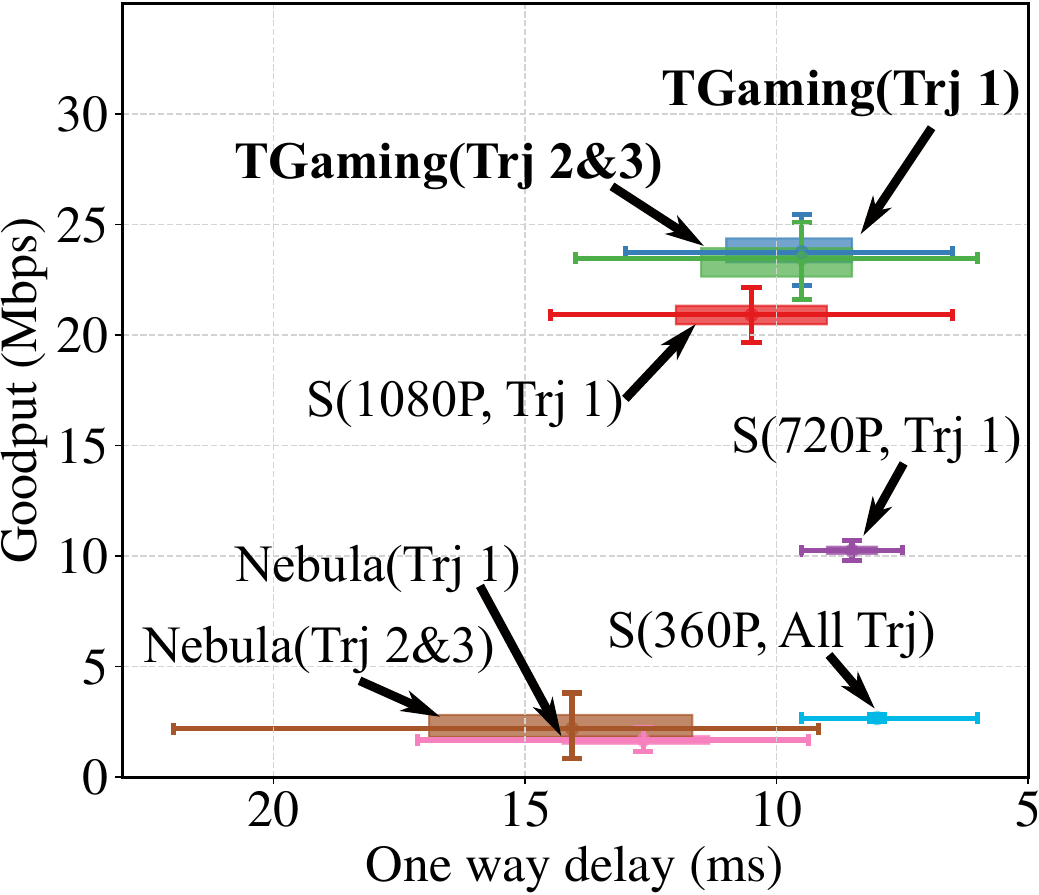}
        \label{fig:bw_packed_moving}}
     \caption{\emph{Network usage comparison:} \sysname{}, Sunshine (S), and Nebula.
     \emph{Boxes:} middle quartiles; \emph{whiskers:} extreme deciles.} 
     \label{fig:bw_eval}
\end{minipage}
% \hspace{0.015\textwidth}
% \begin{minipage}[b]{0.28\textwidth}
%     \centering
%     \subfigure[Less crowded 5G RAN.]{
%         \includegraphics[width=0.95\linewidth]{./figures/qoe_idle_cropped.pdf}
%         \label{fig:qoe_idle}}\\
%         \vfill
%     \subfigure[Crowded 5G RAN.]{
%         \includegraphics[width=0.95\linewidth]{./figures/qoe_packed_cropped.pdf}
%         \label{fig:qoe_packed}}
%      \caption{QoE comparison: whis\-kers indicate 10th\fshyp{}90th percentiles.} \label{fig:qoe}
% \end{minipage}
\end{figure*}

\begin{figure}[tb]
    \centering
    \subfigure[Bit rate estimation for UE 1 (up) and UE 2 (down).]{
        \includegraphics[width=0.4\linewidth]{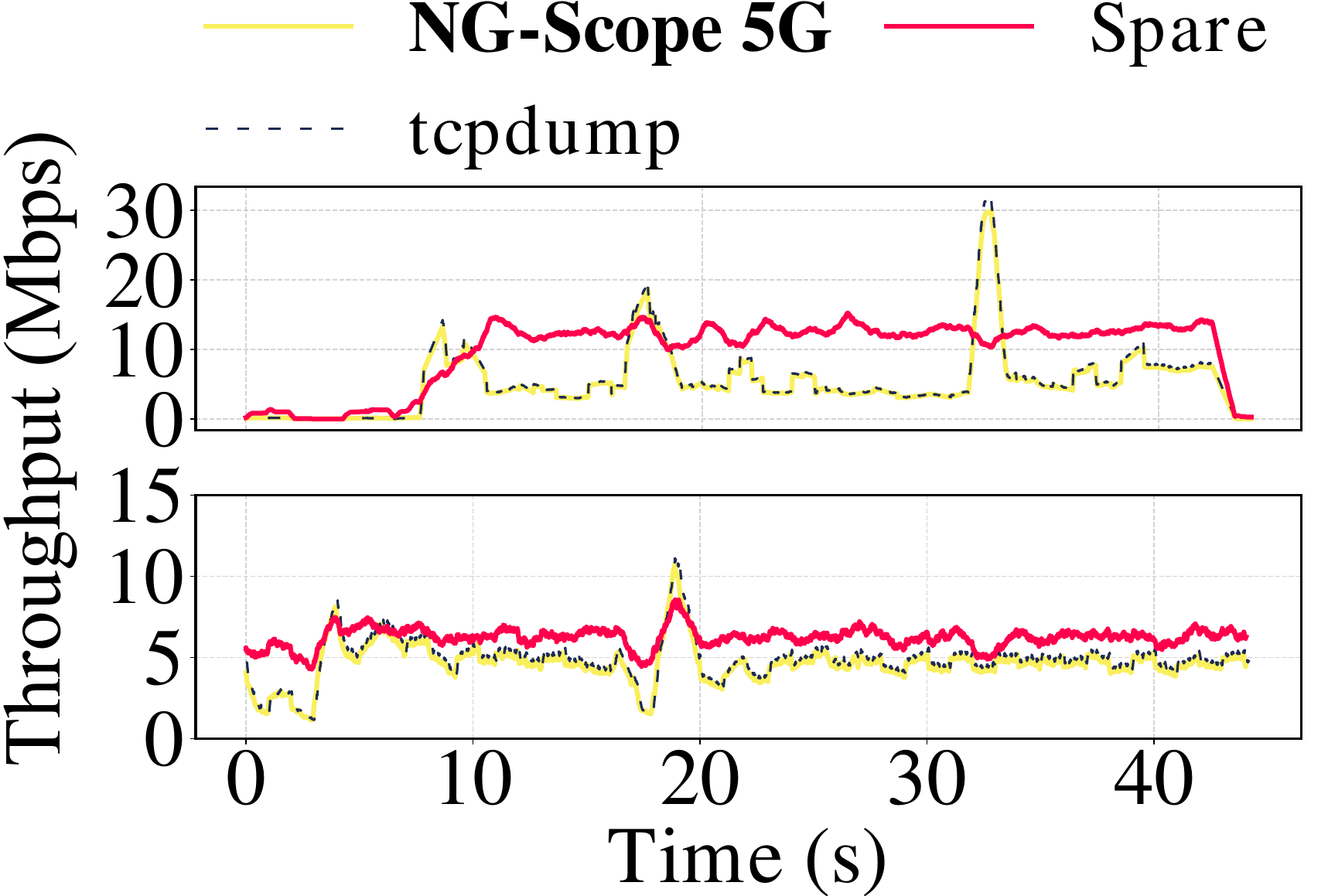}
        \label{fig:demo_bps}}
        \hspace{0.25in}
    \subfigure[Used PRB and fair share PRB estimated by \sysname{} for UE 1 and UE 2.]{
        \includegraphics[width=0.4\linewidth]{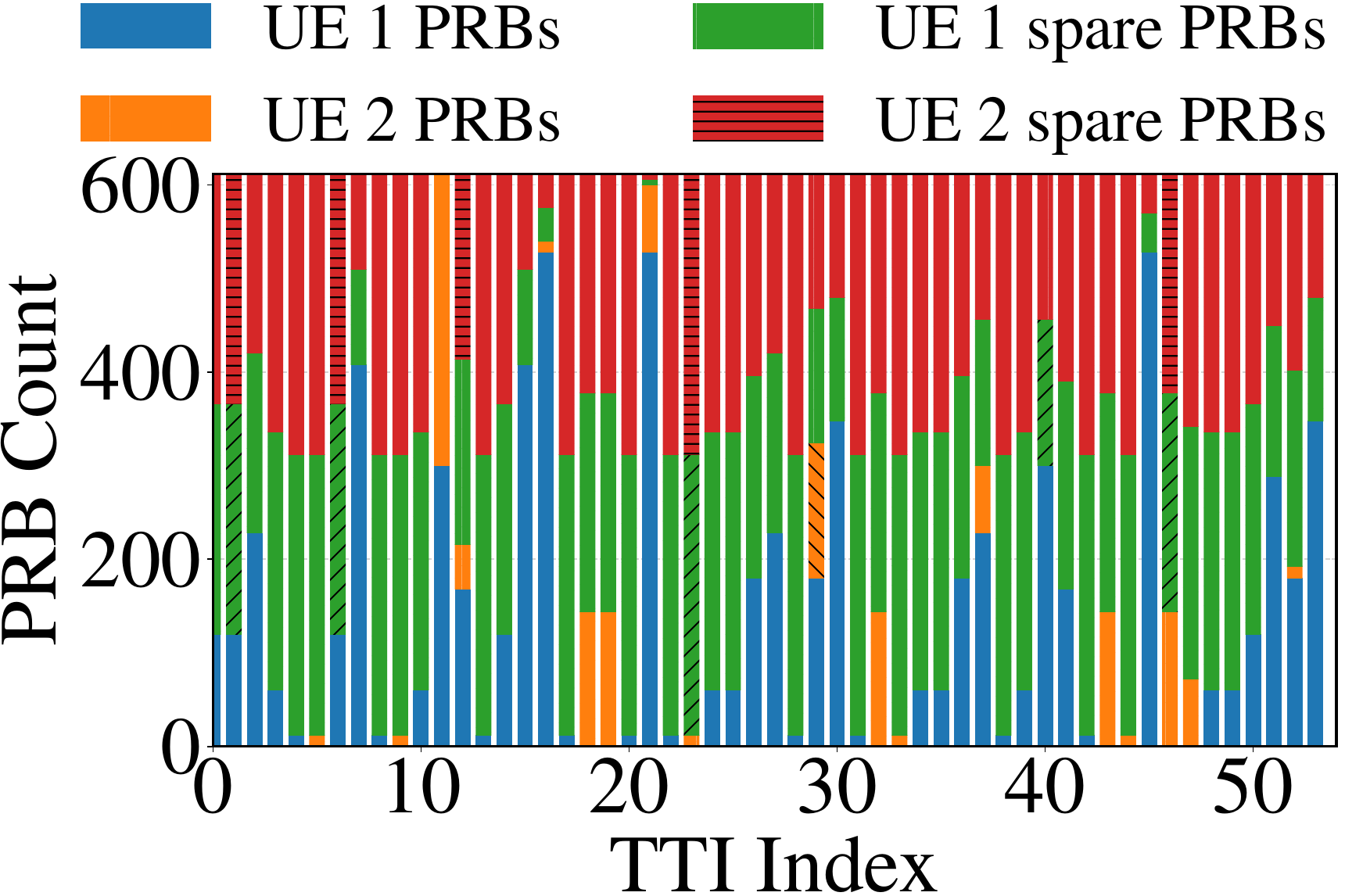}
        \label{fig:demo_prb}}
     \caption{\emph{\toolname{} telemetry example trace:} We calculate the bit rate over time. Then we split the free PRBs equally among UEs and use the same parameters in used PRBs to calculate the bit rate provide by the free PRBs. } \label{fig:demo_2ue}
\end{figure}

\subsubsection{Time series}
We demonstrate \toolname{} telemetry resolution through a
time series shown in \cref{fig:demo_2ue}.
During this demonstration, we connect 2~UEs to the \aethNet{}
network, then use both \toolname{} and tcpdump \cite{tcpdump_official_2023} 
on each phone to calculate each UE's downlink network bandwidth.
UE~1 plays a mobile cloud game through the moonlight mobile client 
 and UE 2 is watching a live stream on Twitch.
As we can see from \cref{fig:demo_bps}, the capacity
estimation for each UE is highly accurate in time (the curve tracks just under 
ground truth).
Turning to \cref{fig:demo_prb} we see \toolname{} calculates
the spare bit rate for each UE through the fair share of spare PRBs.
Even though the number of fair share PRBs are the same for each UE, the 
calculated spare bit rates are different. This is because two UEs have different
MCS indices and coding rates in the same TTI.

\subsubsection{Decoding accuracy.} 
We evaluate \toolname{} telemetry accuracy both via DCI miss rate and 
PRB decoding accuracy.  For this experiment, we use the \srsNet{}
fully open\hyp{}source network.  This enables us to collect 
detailed physical layer \textit{ground truth} for all UEs from srsRAN's log, in 
terms of TTI index, DCI content and downlink grants 
(frequency and time resource allocation, and TBS). 
We match the number of DCIs captured by \toolname{} and srsRAN's log using the 
timestamp and the TTI index, through which we calculate a DCI decoding
\emph{miss rate}.
%During the evaluation, UEs keep the cellular connection through watching Youtube videos or 
%downloading files, so UEs will receive DCIs for both uplink and downlink.
We use the start and end timestamp and TTI from \toolname{} as an \emph{observation period}, 
and from the observation period, we take all the DCIs from srsgNB's log as the ground truth.
We repeat each evaluation five times; each observation period lasts approximately five minutes.

\begin{figure}
    \begin{minipage}{0.4\linewidth}
        \centering
        \includegraphics[width=0.98\textwidth]{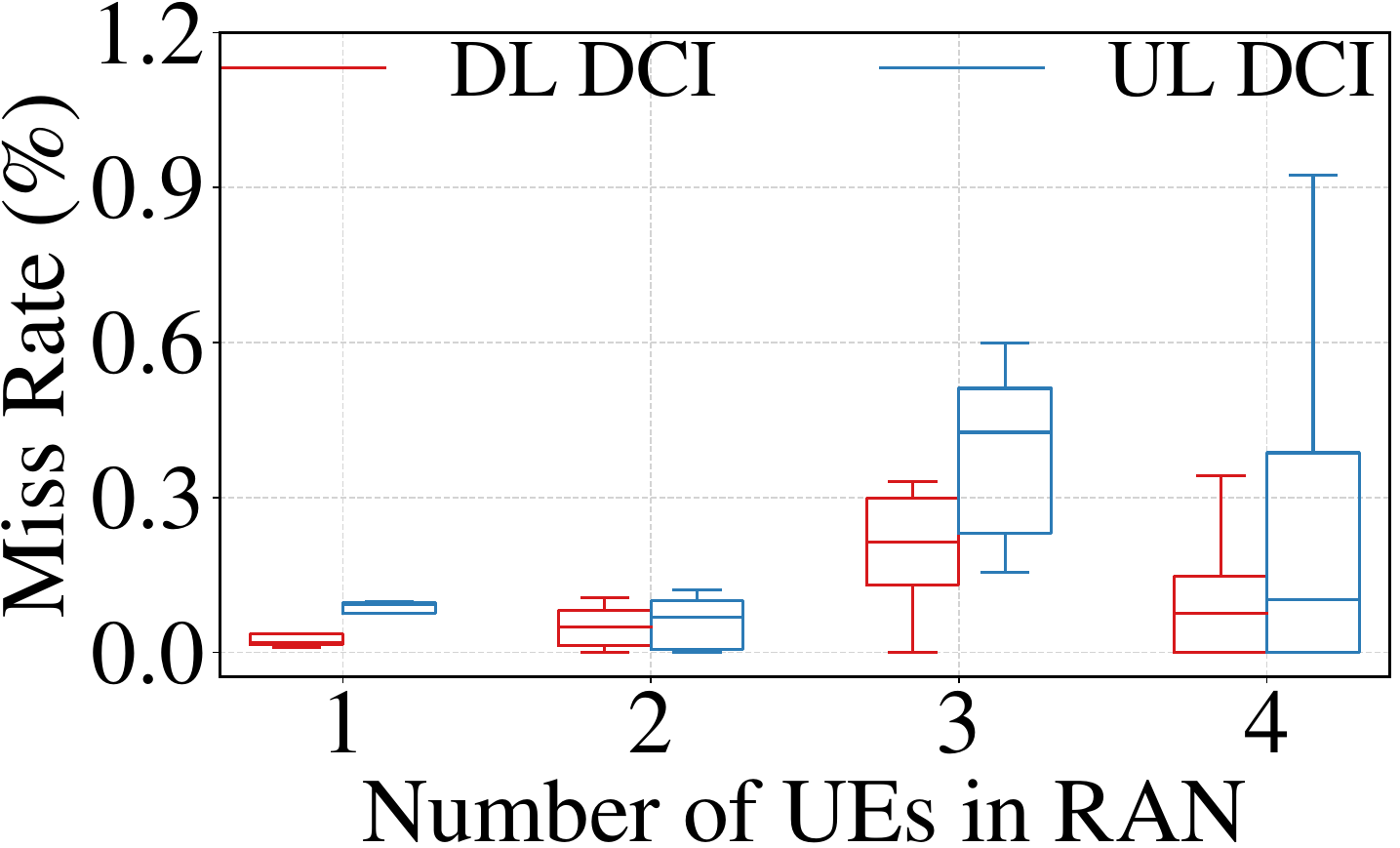}
        \caption{\emph{DCI miss rate} with different number of UEs.}\label{fig:dci_eval_dr}
    \end{minipage}
    \hspace{0.25in}
    % \hfill
    \begin{minipage}{0.4\linewidth}
        \centering
        \includegraphics[width=0.98\textwidth]{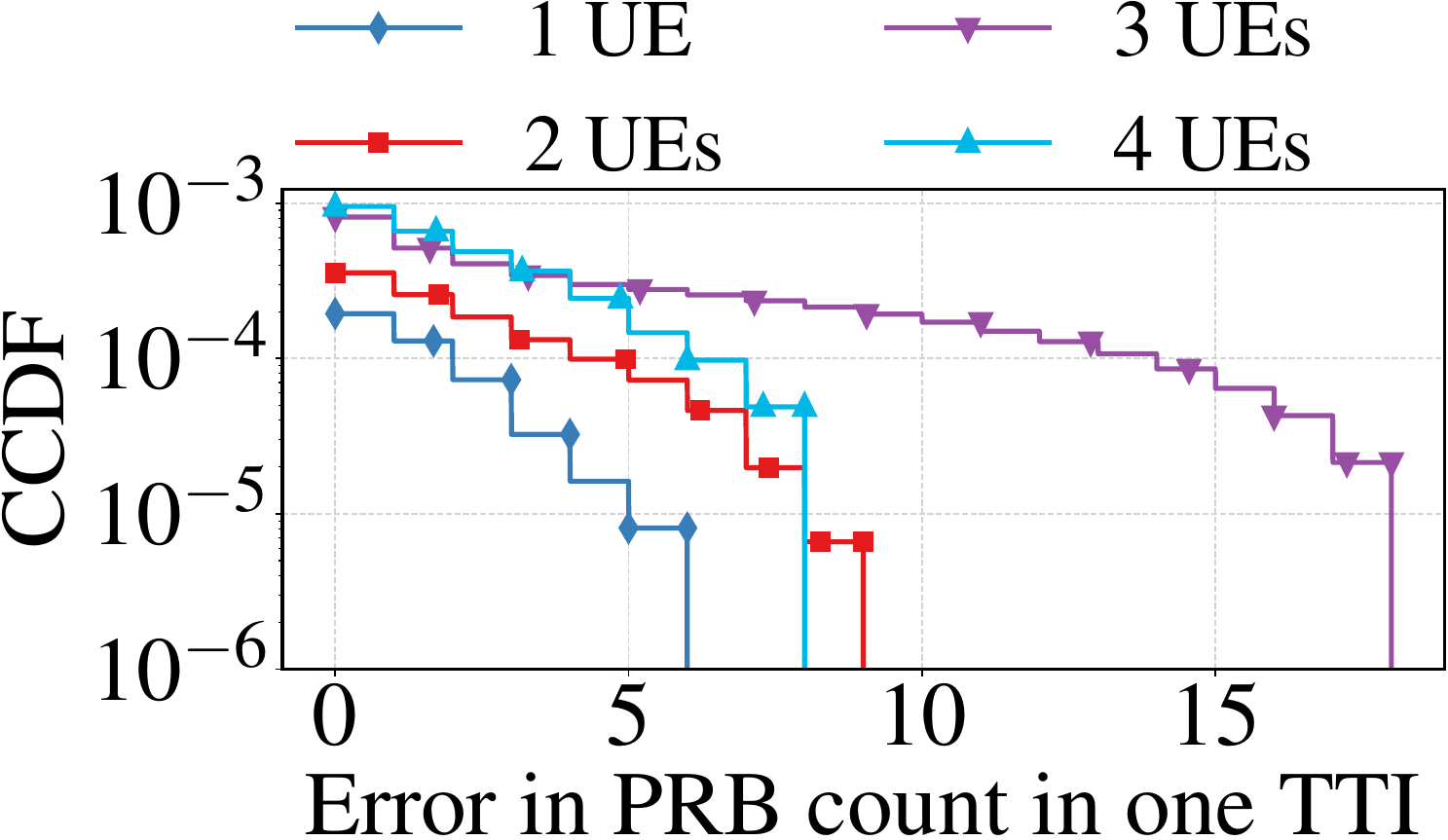}
        \caption{\emph{PRB decoding errors} with different number of UEs.}\label{fig:dci_eval_prb}
    \end{minipage}
    % \begin{minipage}{0.31\textwidth}
    %     \centering
    %     \includegraphics[width=0.95\textwidth]{./figures/dci_eval_dr_cropped.pdf}
    %     \caption{TBS decoding correctness with different UEs in the RAN.}\label{fig:dci_eval_tbs}
    % \end{minipage}\hfill
\end{figure}

Firstly, we test the DCI miss rate for each UE, we connect UEs into the srsgNB, 
and in the mean time use \toolname{} to decode the DCIs.
Then we match the DCIs decoded by \toolname{} and DCIs in srsgNB's log based on the 
timestamp and TTI indexes to calculate the miss rate for each UE and put them 
together to get the distribution.
\toolname{} achieves a very low DCI miss rate---as \cref{fig:dci_eval_dr} shows,
\toolname{} detects the vast majority of downlink and uplink DCIs, with 
miss rates of 0.14\% and 0.22\%, respectively:
two to three 9's of reliability.

\begin{figure*}
    \centering
    \begin{minipage}[b]{\linewidth}
        \centering
        \subfigure[Static UEs.]{
        \includegraphics[width=0.31\linewidth]{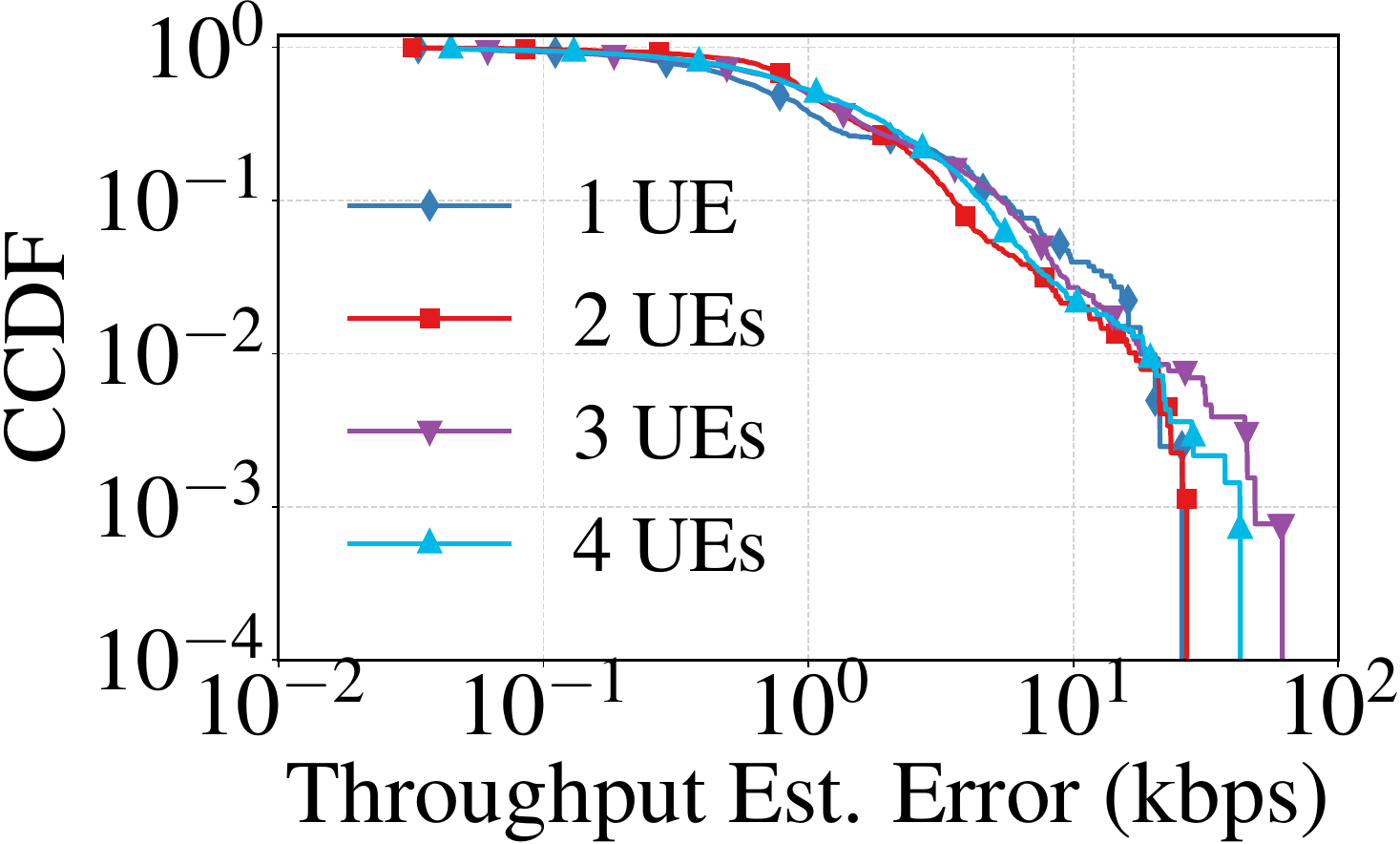}
        \label{fig:bw_est_static}}
        \hfill
    \subfigure[Blocked UEs.]{
        \includegraphics[width=0.31\linewidth]{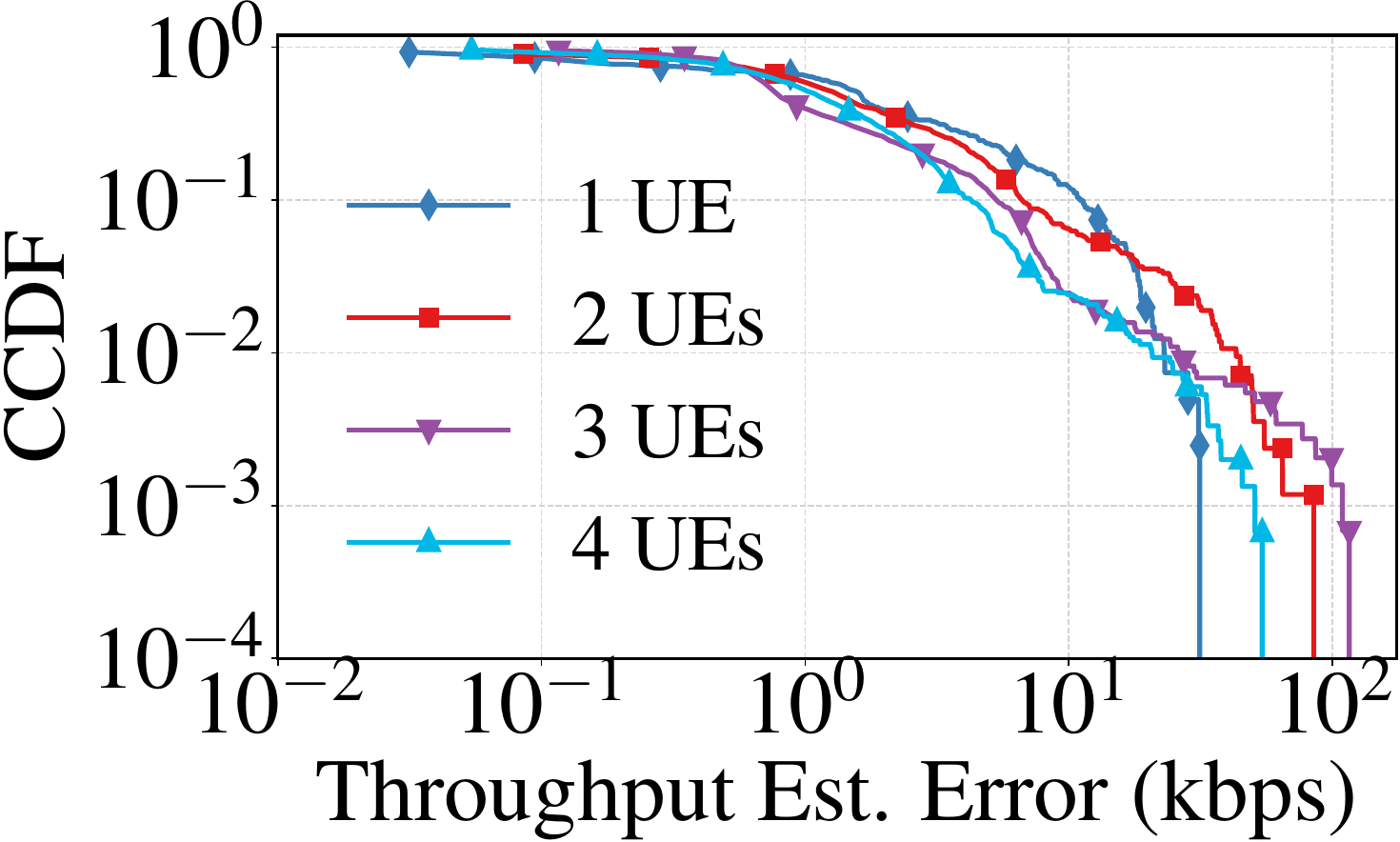}
        \label{fig:bw_est_blockage}}
        \hfill
    \subfigure[Moving UEs.]{
        \includegraphics[width=0.31\linewidth]{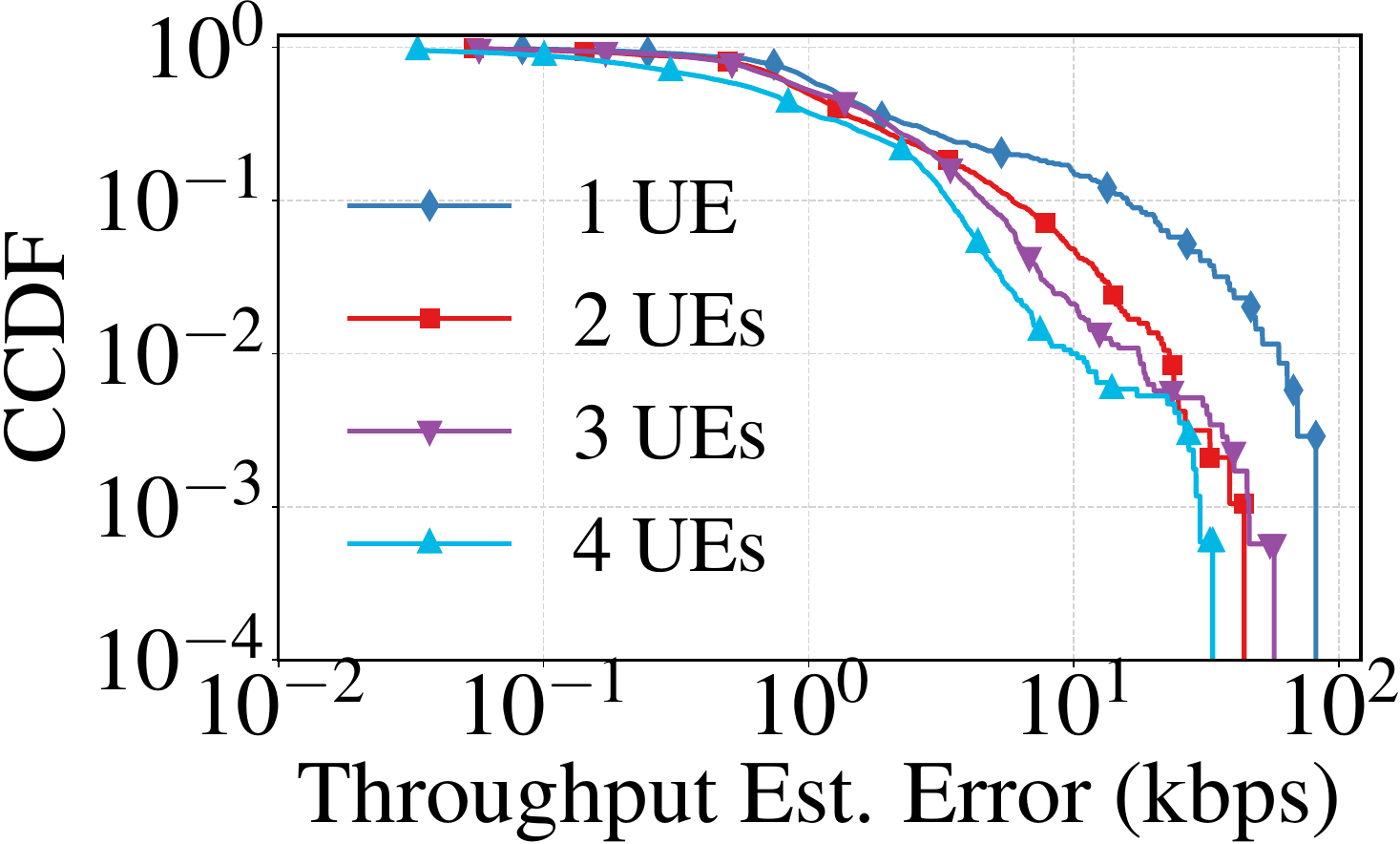}
        \label{fig:bw_est_moving}}
     \caption{\emph{Accuracy of \toolname{} telemetry} throughput estimation (ground truth: tcpdump at the UE.)} \label{fig:bw_est}
    \end{minipage}
\end{figure*}

Secondly, we test PRB decoding correctness in the same setting.
We compare the downlink and uplink grants decoded by \toolname{} and the corresponding 
grants in the srsgNB log.
After we match the DCIs, we compare the decoded number of PRBs with the ground truth 
within each TTI (0.5~ms with 30~kHz subcarrier spacing) and calculate the number 
of errors in the \toolname{}\hyp{}decoded PRB count, per TTI.
\cref{fig:dci_eval_prb} shows that \toolname{} can achieve an average of 0.017 PRB estimation errors 
per TTI, and most of the time, the PRB estimation error is zero.
As we can observe from \cref{fig:demo_prb}, the number 
of PRBs allocated for each UE can be up to several hundreds, and so errors of this 
level will not materially affect the bit rate estimation, as we show next.
This demonstrates that \toolname{} can decode physical layer PRB information from 
the base station with high accuracy and granularity.

% The missed DCI messages and PRBs estimation are mainly caused by the imperfection of RF antennas, where some DCIs are falsely dropped due to low signal SNR.
% %
% We can potentially improve the performance of \sysname{} by using antennas that have the same resonate frequency as the base station.
%
% This evaluation shows that \sysname{} can detect and decode the physical layer information from the base station with high accuracy and great granularity.

\subsubsection{Throughput estimation accuracy} 

Here we evaluate the downlink throughput estimation of \toolname{}, 
in our \aethNet{} network.
In the commercial small cell we do not have access to the physical layer ground truth as 
we did previously, and so for \emph{ground truth} we instead run tcpdump 
\cite{tcpdump_official_2023} on the phone to capture network packets, 
calculate the bit rate and compare the results.
We connect different numbers of UE into the small cell, which use the data 
to watching videos or downloading files, and at the same time we 
use \toolname{} to decode their DCIs.
We test the effectiveness of \toolname{} under different UE usage scenarios, 
including static, moving and blocked, to mimic real\hyp{}life scenarios. 

\toolname{} achieves a highly accurate bandwidth usage estimation:
\cref{fig:bw_est_static} to~\ref{fig:bw_est_moving} show the results:
\toolname{} estimates the UEs' throughput with 75th percentile errors 
of 2.23, 2.53, and 2.24~Kbit\fshyp{}second in static, blocked, 
and moving scenarios, respectively.
The average downlink bit rate for all UEs is 3.35~Mbit\fshyp{}second
in this evaluation, 
so the 75th percentile bit rate estimation errors are under 0.1\%, with 
errors mainly from missed DCIs.

\begin{table}[tb]
%\begin{tabularx}{\linewidth}{*{2}{X}}
%\textbf{DCI Count}  & \textbf{Processing Time ($\mu s$)} \\ \midrule
%1              & $53.74\pm 38.52$     \\
%2              & $70.74 \pm 34.29$    \\
%3              & $97.23\pm 32.56$     \\
%4              & $138.72 \pm 39.40$  \\ \bottomrule
%\end{tabularx}
\begin{tabularx}{\linewidth}{@{}l*{4}{X}@{}}\toprule
\textbf{DCI count:}& 1& 2& 3& 4\\
\textbf{Time ($\boldsymbol{\mu}$s):}& 
    $54 \pm 39$& $71\pm 34$& 
    $97 \pm 33$& $140 \pm 39$ \\\bottomrule
\end{tabularx}

\caption{\emph{Single-thread processing time} per transmission time interval,
for different numbers of DCIs in one TTI.}
\label{tab:proc-time}
\end{table}

\subsubsection{Processing time.} 
Finally, we demonstrate that \toolname{} 
can operate in real time. We evaluate the signal processing time and DCI 
translation time of \toolname{} in one thread with different number of
DCIs in one TTI.
A timely processing is vital for telemetry because we need to decode the DCI within
the time of one TTI, which ranges from 1~ms for 15~kHz subcarrier spacing, 
to 0.25~ms for 60~kHz subcarrier spacing in sub-6~GHz bands.
In our evaluation, the maximum number of DCIs found in one TTI is four, 
including both downlink and uplink, and the
processing time is shown in \cref{tab:proc-time}.
The data show that \toolname{} can process the data efficiently well within 
the time of one TTI.
In this evaluation, we only use one thread of the CPU and so we can easily 
increase processing capacity by exploiting the parallelism of the task.

% \begin{figure}[tb]
%   \centering
%   \includegraphics[width=\linewidth]{figures/eval_setup_cropped.pdf}
%   \caption{UE status setup for cloud gaming evaluations.}
%   \label{fig:eval_locations}
% \end{figure}

%% file: sections/eval_gaming.tex
\subsection{Cloud Gaming Performance}
\label{s:eval:cloudgaming}
\label{s:eval:gaming}

In this part of the evaluation, we run the original and modified cloud gaming host on a 
2-core CPU laptop with 8 GB of memory,
and we run the original and modified moonlight client on 
one of the Motorola phones.  We use our
\aethNet{} Private 5G Standalone network.
All the foregoing machines run Ubuntu Linux 22.04. 

\subsubsection{Methodology}
\label{s:eval:gaming:methodology}

We compare the end-to-end performance of \sysname{} mobile cloud 
gaming head to head against Nebula \cite{alhilal_nebula_2022} and Sunshine.
Sunshine does not adjust bit rate, so we have
enhanced it to run at 1080p, 720p, and 360p,
at 60~fps. 
Nebula (described in \S\ref{s:related}) uses 
forward error correction to adjust its bitrate.  
We configure 
it and \sysname{} to use a maximum resolution of 1080p; 
both use their own ABR algorithms.
Nebula is developed with Python and does not support Android devices.
For a sound comparison, we run Nebula's server code on the same laptop 
that we run Sunshine server on and connect the client to the 5G RAN by 
using USB tethering with the mobile phone.

In our experiments, we collect all the metrics from the mobile cloud 
gaming server and client in real time. 
For \sysname{} and Sunshine, we collect network packets and calculate downlink throughput 
by running tcpdump on the mobile client.
We collect the video's resolution inside the video scheduler and collect 
received frame rate, latency probing results, and packet loss rate on the
client with moonlight built\hyp{}in functions. 
For Nebula, we collect throughput by running tcpdump on the client 
laptop, and we collect resolution, frame rate, and latency probing
results from its built\hyp{}in logs. 
We calculate Nebula's packet loss rate by comparing the actual number of frames
received and the frame index.

We evaluate the performance of mobile cloud gaming using both network metrics and QoE metrics.
Our network metrics contain goodput and one way delay.
We select QoE metrics including video resolution, frame rate, 
network latency, and packet loss rate 
\cite{metzger_comprehensive_2016,laghari_quality_2019} with
overall QoE metric
\begin{equation}
    \mathrm{QoE}\ =\ \alpha R + \beta F - \gamma L - \delta P,
\end{equation}
where $\alpha$ to $\delta$ are the weights for each parameters, $R$ is the resolution, $F$ is the frame rate, $L$ is the network latency and $P$ is the packet loss rate.
We normalize all the parameters into range of $[0,\ 1]$ using min-max normalization, with the global maximal and minimal values in the evaluation.
For resolution, we only use one axis of the screen (width/height) to perform normalization.
We rank the importance of these parameters' importance as: $P > F > L > R$, according to the importance analysis in~\cite{jarschel_gaming_2013,hong_enabling_2015}, and the weights are set as $\alpha=0.5$, $\beta=0.6$, $\gamma=0.7$, and $\delta=0.8$.

To evaluate under rigorous real\hyp{}world scenarios, we measure the performance of 
\sysname{} (T), Sunshine (S), Nebula (N) under different settings: static, 
blocked and moving clients.
\cref{fig:eval_locations} shows the UE's locations and trajectories during this evaluation.
We create two different RAN conditions---a less 
crowded RAN and a crowded RAN. 
For the less crowded RAN, we use other 3 UEs to watch Youtube live stream, 
where there are plenty of spare PRBs for the cloud gaming client.
For the crowded RAN, we use the three other UEs to download big files 
from the Internet using UDP protocol to make the number of spare PRBs close to zero.
For each UE and RAN status, we run each system for five minutes
in static scenarios, and 2--3~minutes in the mobile 
scenarios\footnote{Some Nebula runs are shorter due to crashes, but all
last at least two minutes, and our metrics do not penalize
Nebula in these circumstances.} and repeat this process five times.

% \begin{figure}[tb]
% \centering
%     \subfigure[When 5G RAN is less crowded.]{
%         \includegraphics[width=0.225\textwidth]{./figures/res_idle_cropped.pdf}
%         \label{fig:res_idle}}
%         \hfill
%     \subfigure[When 5G RAN is crowded.]{
%         \includegraphics[width=0.225\textwidth]{./figures/res_packed_cropped.pdf}
%         \label{fig:res_packed}}
%      \caption{Resolution (\textbf{T}: \sysname{}; S: Sunshine; N: Nebula).} \label{fig:res_qoe}
% \end{figure}

\begin{figure}[tb]
\centering
    \includegraphics[width=.7\linewidth]{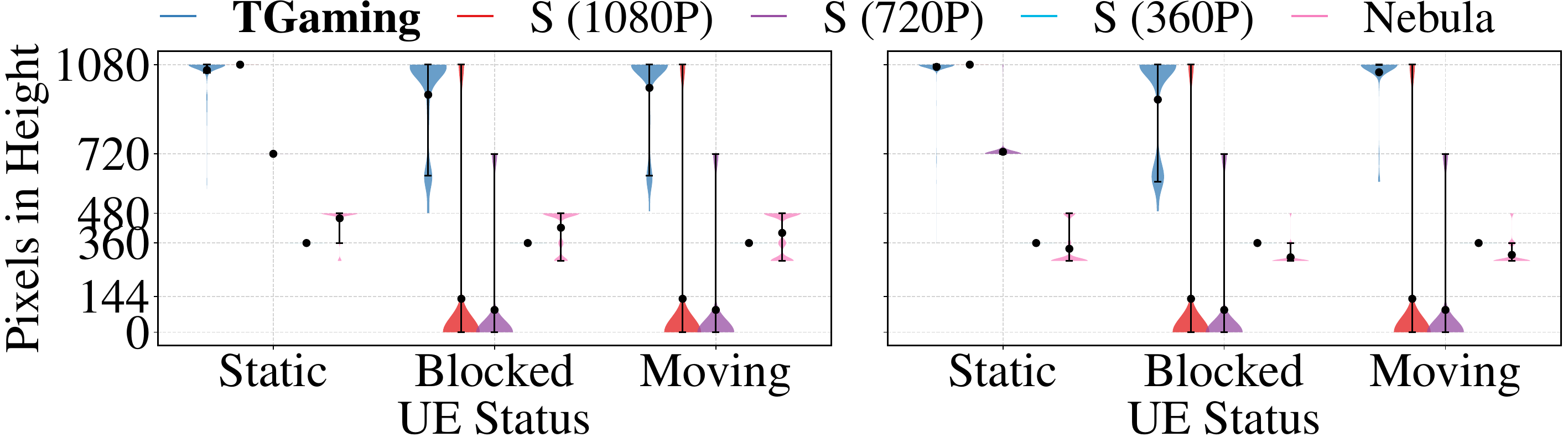}
    \caption{\emph{Video playback resolution} (S: Sunshine). \emph{Left:}~less crowded RAN; 
    \emph{right:}~crowded RAN.\protect\footnotemark} 
    \label{fig:res_qoe}
\end{figure}

\begin{figure}[tb]
\centering
    \includegraphics[width=.7\linewidth]{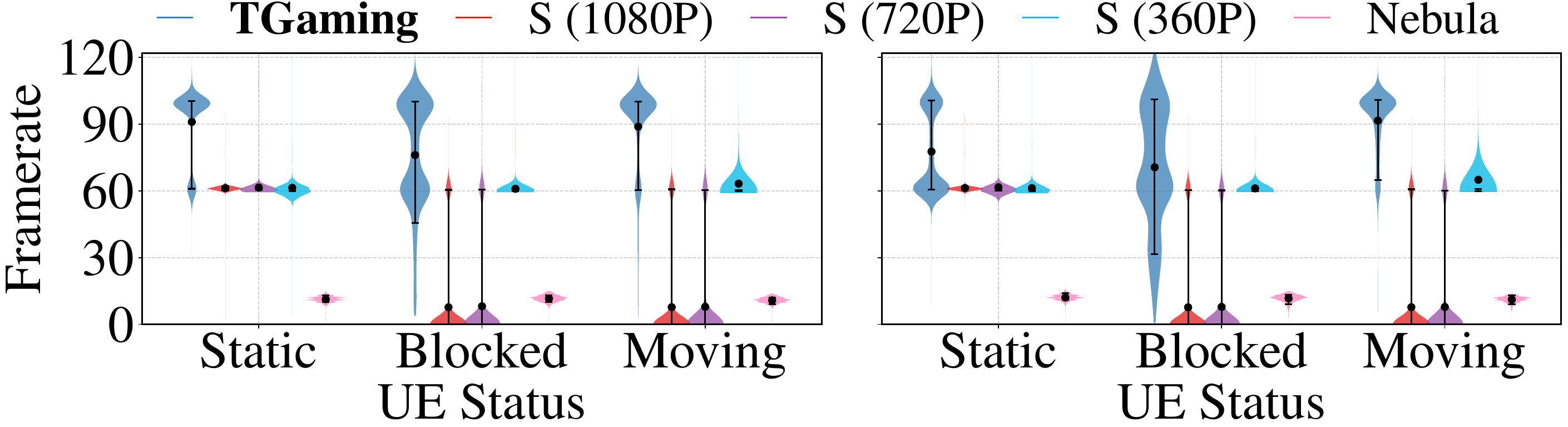}
    \caption{\emph{Video frame rate} in frames\fshyp{}second
     (S: Sunshine).  \emph{Left:}~less crowded RAN; 
    \emph{right:}~crowded RAN.\protect\footnotemark[\value{footnote}]} 
    \label{fig:fps_qoe}
\end{figure}

% \begin{figure}[tb]
% \centering
%     \subfigure[When 5G RAN is less crowded.]{
%         \includegraphics[width=0.225\textwidth]{./figures/fps_idle_cropped.pdf}
%         \label{fig:fps_idle}}
%         \hfill
%     \subfigure[When 5G RAN is crowded.]{
%         \includegraphics[width=0.225\textwidth]{./figures/fps_packed_cropped.pdf}
%         \label{fig:fps_packed}}
%      \caption{} \label{fig:fps_qoe}
% \end{figure}

\subsubsection{QoE Evaluation}
\label{s:eval:qoe}

We evaluate all three systems under the experimental scenarios introduced above.
\cref{fig:qoe,fig:res_qoe,fig:fps_qoe,fig:plr_qoe}, 
show the performance of all three systems in overall QoE and each
individual QoE metric. 
\cref{fig:qoe} shows the overall QoE results, where \sysname{} has the best 
average QoE across all UE and RAN conditions.
\sysname{} improves QoE by 178.3\% and 249.2\%, compared with
Sunshine's average QoE
when the RAN is less crowded and more crowded, relatively.
Sometimes (in blocked Locations~2 and~3 and 
moving trajectories~2 and~3) 
Sunshine stalls due to sending at too high a bit rate.
Nebula has the lowest QoE result because it conservatively uses a 
low resolution and frame rate to maintain the seamless experience.
When the UE is blocked or moving to the area where the signal is weak 
(blocked and moving UE scenarios), \sysname{} actively changes the resolution
and frame rate, yielding a high variance in resolution and frame 
rate (in \cref{fig:res_qoe,fig:fps_qoe}).
Nebula uses a low resolution and low frame rate video in these conditions, 
which contributes to its low overall QoE results.
As for packet loss rate, all three system have a low packet loss rate when
the UE is static (\cref{fig:plr_qoe}), \sysname{} and Nebula has
a slightly worse packet loss 
rate when the RAN is crowded.  
This evaluation shows that \sysname{} can improve the QoE when the RAN 
connection is ideal and help to overcome packet loss
caused by a weak signal.

% \begin{figure}
% \centering
%     \subfigure[When 5G RAN is less crowded.]{
%         \includegraphics[width=0.225\textwidth]{./figures/latency_idle_cropped.pdf}
%         \label{fig:latency_idle}}
%         \hfill
%     \subfigure[When 5G RAN is crowded.]{
%         \includegraphics[width=0.225\textwidth]{./figures/latency_packed_cropped.pdf}
%         \label{fig:latency_packed}}
%      \caption{Latency, where the two ends of the error bar gives the 10th and 90th percentiles {\color{red} duplicated information with \cref{fig:bw_eval}}.} \label{fig:latency_qoe}
% \end{figure}

\subsubsection{Network evaluation}
We begin with an examination of the network usage properties
of all three systems. 
We observe that Sunshine suffers outages
in Loc.~2 and~3 in the blocked scenario,
and Traj.~2 and~3 in the moving 
scenario, where its packet loss rate reaches 100\%, while both 
\sysname{} and Nebula can work under these circumstances.
Thus, we exhibit the results separately for these challenging environments.
\cref{fig:bw_eval} shows the results: when the RAN is 
less crowded, \sysname{} achieves high throughput and 
low latency when channel quality is 
good (\cref{fig:bw_idle_static}, Loc.~1 in \cref{fig:bw_idle_blockage} 
and Traj.~1 in \cref{fig:bw_idle_moving}).
Otherwise, (Loc.~2 and~3 in \cref{fig:bw_idle_blockage} 
and Traj.~2 and~3 in \cref{fig:bw_idle_moving}), \sysname{} 
matches its bit rate to the RAN's bit rate, which causes 
high throughput and latency variance.  
Sunshine maintains stable performance when the channel is 
good, but cannot adjust itself to accommodate a poor channel 
so loses its functionality.
Nebula is conservative, keeping resolution and 
frame rate low to keep running in a poor network.
When the RAN is crowded, \sysname{} increases its bit rate 
with the cost of higher variance in latency if the physical 
channel quality is good.
In a weak channel, (Loc.~2 and~3 in 
\cref{fig:bw_packed_blockage} and Traj.~2 and 3 in 
\cref{fig:bw_packed_moving}), both \sysname{} and Nebula provide a 
seamless experience, but Nebula makes a conservative bandwidth 
choices while \sysname{} better utilizes spare RAN capacity while
maintaining low latency.

\begin{figure}[tb]
\centering
    \includegraphics[width=.7\linewidth]{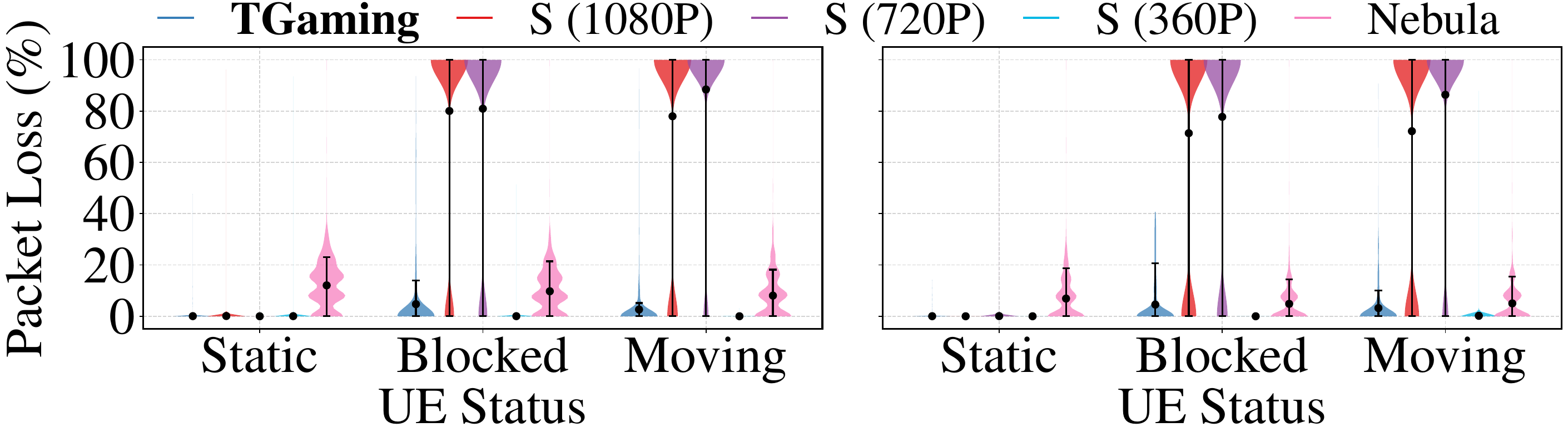}
     \caption{\emph{Application\hyp{}level packet loss rate} (S: Sunshine).  
     \emph{Left:}~less crowded RAN; 
    \emph{right:}~crowded RAN.\protect\footnotemark[\value{footnote}]} \label{fig:plr_qoe}
\end{figure}

\footnotetext{In \cref{fig:res_qoe,fig:fps_qoe,fig:plr_qoe}, the data series order
matches the legend order.}

%% file: sections/concl.tex
\section{Conclusion}
\label{s:concl}

This work has presented the first open\hyp{}source 5G Standalone 
RAN passive network 
telemetry tool, and a telemetry architecture for network applications
and end\hyp{}to\hyp{}end transport layer protocols
that sheds light into the vagaries of the wireless RAN so that they may better adapt
their behavior. The \toolname{} telemetry tool does 
not require the cooperation of mobile network operators, chipset manufacturers,
or phone manufacturers, nor does it require modifications to the UE.  
Our mobile cloud gaming 
implementation and evaluation has demonstrated significant
QoE improvements over leading open source game streaming platforms, 
in a variety of static
and mobile settings.
%We have shown that moonlight's bit rate estimation function as the scheduling guidance.
%Even so, the real-time bit rate of mobile cloud gaming's can vary in different situations, \textit{e.g.} high speed moving scenery can have $2\times$ as much bit rate as static scenery according to our observation.
%Unlike the video on\hyp{}demand task that has known pre\hyp{}rendered video chunks and bit rate requirements, an accurate real-time cloud gaming bit rate predictor is vital for cloud gaming video scheduler.
%In both LTE and 5G, 
%carrier aggregation allows one UE to connect to 
%multiple cells simultaneously to improve 
%throughput \cite{xie_pbe-cc_2020,xie_ng-scope_2022}.
%Here we don't test \sysname{} in such circumstances,
%but can achieve the same functionality 
%by decoding DCIs from multiple 
%cells with multiple radios, and matching these DCIs through TTI indexes.
%We will continue working on \sysname{} to make it a fully
%3GPP\hyp{}standard featured tool for 5G SA telemetry.
Here we use 5G SA telemetry 
information to schedule downlink traffic (cloud gaming video), 
but \toolname{} can also uplink telemetry information 
accurately, which can help,
%to optimize 
%uplink transmission where the uplink throughput and 
%latency are also critical, 
\textit{e.g.} web conference 
and outdoor live streams.  
%\textbf{3)~URLLC:} our 5G base stations do not support
%URLLC, which is especially relevant to the control uplink
%\textbf{3)~X:} a 
%\textbf{3)~RRC reconfiguration:} 
%After random access, the gNB may send a further message to 
%the UE to further specify or update the RAN parameters specified 
%in MSG 4. This too is short\hyp{}term future work.
\paragraph{Ethics statement} Cellular telemetry is broadcasted without encryption, so is accessible to anyone with easily-obtainable equipment, and is used here for the sole purpose of communications network design.  Telemetry data itself does not identify persons, and the ability to combine the data with another dataset to identify an individual is so attenuated as to be generally impracticable.  The relevant IRB has waived the associated master study.

%\paragraph{Open source statement.}  Pending acceptance the authors will release 
%their telemetry tool to the community.

%% file: sections/appendix.tex
\section{Transport Block Size Calculation}
\label{appd:tbs}

The following development restates the 3GPP standard TBS calculation
\cite{3gpp_release_2022} for convenience here.

Firstly, we calculate the intermediate variable $N_{RE}$:
\begin{eqnarray}
N_{RE}^{'} &=& N_{sc}^{RB} \times N_{symb}^{sh} - N_{DMRS}^{PRB} - N_{oh}^{PRB},\\
N_{RE} &=& min(156, N_{RE}^{'})*n_{PRB}.
\end{eqnarray}
$N_{sc}^{RB}$ is the number of subcarriers per resource block and it's 12.
$N_{symb}^{sh}$ is the number of time domain OFDM symbols allocated through DCI, which can be found in $t\_alloc$ in the DCI grant in Appendix~\ref{appd:dci-and-grant}.
$N_{DMRS}^{PRB}$ is the number of resource elements for DMRS per PRB and $N_{oh}^{PRB}$ is the overhead of PDSCH, which are both defined in RRC messages.
$n_{PRB}$ is the frequency domain resource block allocated by DCI, which can be found in $f\_alloc$ in the DCI grant in Appendix~\ref{appd:dci-and-grant}.
$N_{RE}$ represents the effective resource elements (1 OFDM symbol $\times$ 1 subcarrier) allocated for the UE in the DCI.

The second step is to calculate $N_{info}$:
\begin{eqnarray}
N_{info} = N_{RE} \times R \times Q_{m} \times v.
\end{eqnarray}
$R$ is the code rate and $Q_m$ is the modulation order, which are delivered through the DCI's MSC value and the UE checks the predefined tables with it \cite{3gpp_release_2022}.
And $v$ is number of layers, transferred in the MSG 4's element "$pdsch-ServingCellConfig$": "$maxMIMO-Layers$".

The final step is to calculate the TBS, which contains some value determination:
\begin{itemize}
    \item If $N_{info} \le 3824$, we have $N_{info}^{'} = 2^{n} \times round(\frac{N_{info} - 24}{2^n})$, where $n=\lfloor log_2(N_{info} -24)\rfloor-5$.
    Then, if $R \le 1/4$, we have $TBS = 8C \lceil \frac{N^{'}_{info} + 24C}{8C} \rceil - 24$, where $C = \lceil \frac{N^{'}_{info} + 24}{3814} \rceil$.
    If $R > 1/4$, we have $TBS = 8\lceil \frac{N^{'}_{info} + 24}{8} \rceil - 24$ if $N^{'}_{info} < 8424$, and $TBS = 8C \lceil \frac{N^{'}_{info} + 24C}{8C} \rceil - 24$ if $N^{'}_{info} \ge 8424$, where $C = \lceil \frac{N^{'}_{info} + 24}{8424} \rceil$.
    \item If $N_{info} > 3824$, we have $N_{info}^{'} = max(24, 2^n\lfloor \frac{N_{info}}{2^n} \rfloor)$, where $n = max(3, \lfloor log_2 (N_{info})\rfloor - 6)$.
    Then we check table 5.1.3.2-2 in \cite{3gpp_release_2022} to determine the TBS.
\end{itemize}

\section{RTSP Introduction} 
\label{appd:rtsp}

\begin{figure}[tb]
  \centering
  \includegraphics[width=\linewidth]{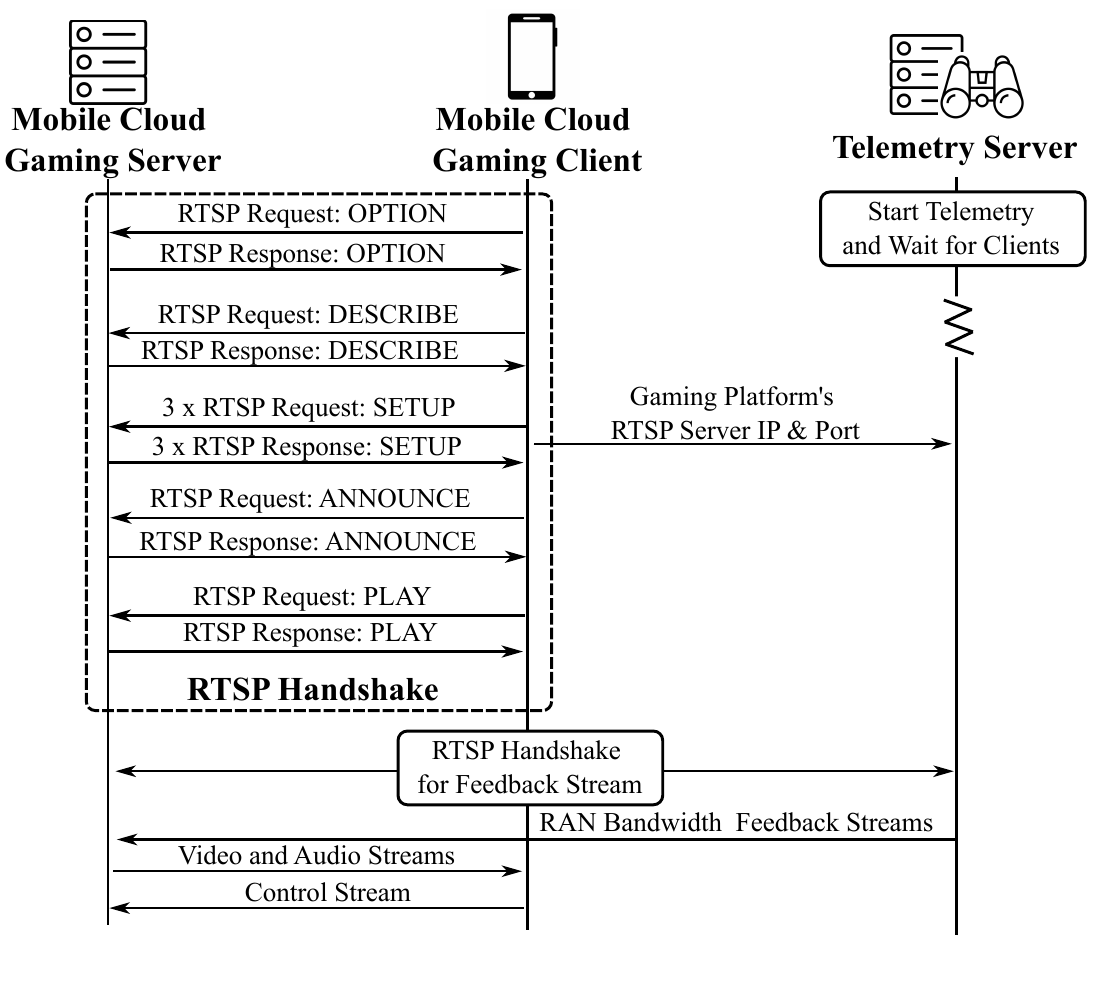}
  \caption{RTSP initiated data flow between mobile cloud gaming server, client and telemetry server. }
  \label{fig:rtsp-design}
\end{figure}

RTSP is an application layer protocol that the server and client use to set up real-time streams.
The detailed RTSP handshake message flow is shown in \cref{fig:rtsp-design}, the client sends request messages and the server replies the corresponding information.
The client first sends "OPTION" request to ask other RTSP messages that the RTSP server supports, in sunshine, the other supported RTSP messages are "DESCRIBE", "SETUP", "ANNOUNC", and "PLAY", as in \cref{fig:rtsp-design}.
Then, the client sends "DESCRIBE" message to ask the streams that the server provides, where in the gaming platform are downlink video and audio streams and uplink user input stream.
With these information, the client asks about the ports of these streams it wants to connects through multiple "SETUP" messages.
When the client is ready to connect to these streams, it tells the gaming server through "ANNOUNCE" message.
Then the client sends the "PLAY" message to actually start the gaming service.

After the gaming client tells the telemetry server the IP and port of the gaming server, the telemetry server can use the similar messages exchange ("OPTION" to "PLAY") to connect to the server's feedback stream.
Since the RTSP server is implemented in a event-driven way, the telemetry server's RTSP handshake can happen anytime even in the middle of client's RTSP handshake, not strictly ordered as in \cref{fig:rtsp-design}.

\section{Sample of RRC Message}~\label{appd:rrc}
We list some key components in SIB 1 and MSG 4 here.

\subsection{SIB 1 Sample}~\label{appd:sib1}
The key components for telemetry in SIB 1 that we decoded from the Sercomm small cell are shown as below:
\begin{footnotesize}
\begin{verbatim}
"SIB1": {
  ...
  "servingCellConfigCommon": {
    "downlinkConfigCommon": {
      "frequencyInfoDL": {
        "frequencyBandList": [
          {
            "freqBandIndicatorNR": 48
          }
        ],
        "offsetToPointA": 24,
        "scs-SpecificCarrierList": [
          {
            "offsetToCarrier": 0,
            "subcarrierSpacing": "kHz30",
            "carrierBandwidth": 51
          }
        ]
      },
      "initialDownlinkBWP": {
        "genericParameters": {
          "locationAndBandwidth": 13750,
          "subcarrierSpacing": "kHz30"
        },
        "pdcch-ConfigCommon": {
          {
            "controlResourceSetZero": 10,
            "searchSpaceZero": 0,
            "commonSearchSpaceList": [
              {
                "searchSpaceId": 1,
                "controlResourceSetId": 0,
                "monitoringSlotPeriodicityAndOffset": {
                },
                "monitoringSymbolsWithinSlot": 
                  "10000000000000",
                "nrofCandidates": {
                  "aggregationLevel1": "n0",
                  "aggregationLevel2": "n0",
                  "aggregationLevel4": "n2",
                  "aggregationLevel8": "n0",
                  "aggregationLevel16": "n0"
                },
                "searchSpaceType": {
                  "common": {
                    "dci-Format0-0-AndFormat1-0": {
                    }
                  }
                }
              }
            ],
            "searchSpaceSIB1": 0,
            "searchSpaceOtherSystemInformation": 1,
            "pagingSearchSpace": 1,
            "ra-SearchSpace": 1
          }
        },
        "pdsch-ConfigCommon": {
          {
            "pdsch-TimeDomainAllocationList": [
              {
                "k0": 0,
                "mappingType": "typeA",
                "startSymbolAndLength": 53
              },
              {
                "k0": 0,
                "mappingType": "typeA",
                "startSymbolAndLength": 67
              }
            ]
          }
        }
      },
      "bcch-Config": {
        "modificationPeriodCoeff": "n2"
      },
      "pcch-Config": {
        "defaultPagingCycle": "rf64",
        "nAndPagingFrameOffset": {
        },
        "ns": "one",
        "firstPDCCH-MonitoringOccasionOfPO": {
          "sCS30KHZoneT-SCS15KHZhalfT": [
            1
          ]
        }
      }
    },
    ...
}.
\end{verbatim}
\end{footnotesize}

\subsection{MSG 4 Sample}~\label{appd:msg4}
The key components for telemetry in MSG 4 that we decoded from the Sercomm small cell are shown as follow:

\begin{footnotesize}
\begin{verbatim}
"RRCSetup": {
    ...
    "masterCellGroup": {    
      ...
      "physicalCellGroupConfig": {
        "p-NR-FR1": 26,
        "pdsch-HARQ-ACK-Codebook": "dynamic"
      },
      "spCellConfig": {
        "rlmInSyncOutOfSyncThreshold": "n1",
        "spCellConfigDedicated": {
          "initialDownlinkBWP": {
            "pdcch-Config": {
              {
                "controlResourceSetToAddModList": [
                  {
                    "controlResourceSetId": 1,
                    "frequencyDomainResources": 
                      "111111110000000000000...",
                    "duration": 2,
                    "cce-REG-MappingType": {
                    },
                    "precoderGranularity": "allContiguousRBs",
                    "pdcch-DMRS-ScramblingID": 5
                  }
                ],
                "searchSpacesToAddModList": [
                  {
                    "searchSpaceId": 2,
                    "controlResourceSetId": 1,
                    "monitoringSlotPeriodicityAndOffset": {
                    },
                    "monitoringSymbolsWithinSlot": 
                      "10000000000000",
                    "nrofCandidates": {
                      "aggregationLevel1": "n4",
                      "aggregationLevel2": "n4",
                      "aggregationLevel4": "n2",
                      "aggregationLevel8": "n1",
                      "aggregationLevel16": "n0"
                    },
                    "searchSpaceType": {
                      "ue-Specific": {
                        "dci-Formats": "formats0-1-And-1-1"
                      }
                    }
                  }
                ]
              }
            },
            "pdsch-Config": {
              {
                "dataScramblingIdentityPDSCH": 5,
                "dmrs-DownlinkForPDSCH-MappingTypeA": {
                  {
                    "dmrs-AdditionalPosition": "pos1"
                  }
                },
                "resourceAllocation": 
                  "resourceAllocationType1",
                "pdsch-TimeDomainAllocationList": {
                  [
                    {
                      "k0": 0,
                      "mappingType": "typeA",
                      "startSymbolAndLength": 53
                    },
                    {
                      "k0": 0,
                      "mappingType": "typeA",
                      "startSymbolAndLength": 67
                    }
                  ]
                },
                "rbg-Size": "config1",
                "maxNrofCodeWordsScheduledByDCI": "n1",
                "prb-BundlingType": {
                  "staticBundling": {
                    "bundleSize": "wideband"
                  }
                }
              }
            }
          },
          "firstActiveDownlinkBWP-Id": 0,
          "defaultDownlinkBWP-Id": 0,
          "uplinkConfig": {
            "initialUplinkBWP": {
              "pucch-Config": {
                {
                    ...
                }
              },
              "pusch-Config": {
                {
                  "dataScramblingIdentityPUSCH": 5,
                  "dmrs-UplinkForPUSCH-MappingTypeA": {
                    {
                      "dmrs-AdditionalPosition": "pos1",
                      "transformPrecodingDisabled": {
                        "scramblingID0": 5
                      }
                    }
                  },
                  "pusch-PowerControl": {
                    "tpc-Accumulation": "disabled",
                    "msg3-Alpha": "alpha1",
                    "p0-AlphaSets": [
                      {
                        "p0-PUSCH-AlphaSetId": 0,
                        "p0": 0,
                        "alpha": "alpha1"
                      }
                    ],
                    "deltaMCS": "enabled"
                  },
                  "resourceAllocation": 
                    "resourceAllocationType1",
                  "pusch-TimeDomainAllocationList": {
                    [
                      {
                        "k2": 4,
                        "mappingType": "typeA",
                        "startSymbolAndLength": 55
                      },
                      ...
                    ]
                  },
                  "transformPrecoder": "disabled",
                  ...
                }
              }
            },
            "firstActiveUplinkBWP-Id": 0
          },
          "pdsch-ServingCellConfig": {
            {
              "nrofHARQ-ProcessesForPDSCH": "n16",
              "maxMIMO-Layers": 2
            }
          },
          "tag-Id": 0
        }
      }
    }
}
\end{verbatim}
\end{footnotesize}

\subsection{DCI and Grant}~\label{appd:dci-and-grant}
A downlink DCI looks like:
\begin{footnotesize}
\begin{verbatim}
DCI:
    c-rnti=0x4296, 
    dci=1_1,
    ss=ue,
    L=0,
    cce=7,
    f_alloc=0x33,
    t_alloc=0x0,
    mcs=27,
    ndi=0,
    rv=0,
    harq_id=11,
    dai=2,
    tpc=1,
    harq_feedback=2,
    ports=7,
    srs_request=0,
    dmrs_id=0
\end{verbatim}
\end{footnotesize}

And its translated grant for PDSCH is like:
\begin{footnotesize}
\begin{verbatim}
DMRS:
    type=1
    add_pos=2
    len=single
    typeA_pos=2
    rvd_pattern: 
        begin=0 
        end=275 
        stride=1 
        sc=111111111111 
        symb=00100001000100
Grant:
    rnti=0x4296
    rnti_type=C-RNTI
    k=0
    mapping=A
    t_alloc=2:12
    f_alloc=0:2
    nof_dmrs_cdm_grps=2
    beta_dmrs=1.412538
    nof_layers=2
    n_scid=0
    tb_scaling_field=0
    CW0:
        mod=256QAM
        nof_layers=2
        mcs=27
        tbs=3240
        R=0.926
        rv=0
        ndi=0
        nof_re=432
        nof_bits=3456
SCH:
    mcs_table=256qam
    xoverhead=0
\end{verbatim}
\end{footnotesize}